\documentclass[structabstract]{aa}
\usepackage{graphicx}
\usepackage{txfonts}
\usepackage[]{natbib}
\usepackage{lscape}
\begin{document}
   \title{Normal galaxies in the \emph{XMM-Newton} fields}

   \subtitle{X-rays as a star formation indicator}

   \authorrunning{Rovilos et al.}

   \author{E. Rovilos,\inst{1,2,3} 
           I. Georgantopoulos,\inst{2}
           P. Tzanavaris,\inst{4,5,2}
           M. Pracy,\inst{6}
           M. Whiting,\inst{7}
           D. Woods,\inst{8,9}
           \and
           C. Goudis\inst{2,3}
          }

   \offprints{E. Rovilos\\ \email{erovilos@mpe.mpg.de}}

   \institute{Max Planck Institut f\"{u}r extraterrestrische Physik,
              Giessenbachstra\ss e, 85748, Garching, Germany
         \and
             Institute for Astronomy and Astrophysics, National
             Observatory of Athens, I. Metaxa \& V. Pavlou str, Palaia
             Penteli, 15236, Greece
         \and
             Astronomical Laboratory, Department of Physics, University
             of Patras, 26500, Rio-Patras, Greece
         \and
             Laboratory for X-ray Astrophysics, NASA Goddard Space Flight
             Center, Mail Code 662, Greenbelt, MD 20771, USA
         \and
             Department of Physics and Astronomy, The Johns Hopkins University,
             366 Bloomberg Center, 3400 N. Charles street, Baltimore, MD 21218,
             USA
         \and
             Research School of Astronomy and Astrophysics, Mount Stromlo
             Observatory, Cotter Rd, Weston, ACT 2611, Australia
         \and
             Australia Telescope National Facility, PO Box 76, Epping,
             NSW 1710, Australia
         \and
             School of Physics, University of New South Wales, Sydney NSW 2052,
             Australia
         \and
             Dept. of Physics \& Astronomy, University of British Columbia, 6224
             Agricultural Road, Vancouver B.C., V6T 1Z1, Canada
             }

   \date{Received date; accepted date}

\abstract {We use the first XMM serendipitous source catalogue (1XMM) to
           compile a sample of normal X-ray galaxies}
          {We seek to expand the database of X-ray selected normal galaxies
           at intermediate redshifts and examine the relation between X-ray
           emission and star formation for late-type systems}
          {The candidates are selected based on their X-ray (soft spectra),
           X-ray to optical $(\log(f_{\rm x}/f_{\rm o})<-2)$ and optical
           (extended sources) properties. 44 candidates are found and 35 are
           spectroscopically observed with the Australian National University's
           2.3\,m telescope to examine their nature.}
          {Of the 35 sources observed, 2 are AGN, 11 emission line galaxies, 12
           absorption line galaxies, 6 have featureless spectra while 4 are
           associated with Galactic stars. We combine our emission line sample
           with earlier works forming the most comprehensive X-ray selected galaxy
           sample for the study of the X-ray luminosity to the H$\alpha$
           luminosity - a well-calibrated star-formation indicator - relation.}
          {We find that the X-ray luminosity strongly correlates with the
           H$\alpha$ luminosity, suggesting that the X-rays efficiently
           trace the star-formation.}

   \maketitle
%

\section{Introduction}

X-ray emission from normal galaxies (i.e. galaxies which do not host an AGN)
has been targeted by X-ray telescopes since the early years of X-ray astronomy.
Normal galaxies are
generally divided into two broad morphological categories, late-type (spirals
and irregulars) and early-type (S0 and ellipticals). Early studies of
optically selected samples of early-type galaxies with EINSTEIN
\citep{Trinchieri1985,Fabbiano1987} have examined the relation between the
X-ray, the blue optical and radio luminosities and argued that
the X-ray luminosity arises from a combination of low-mass X-ray binaries
(LMXRBs) and hot gas \citep[e.g.][]{Forman1979}. On the other hand, the X-ray study
of late-type systems with EINSTEIN \citep*{Fabbiano1985,Fabbiano1988} has
revealed a different kind of X-ray correlation for the radio (shallower) and
blue (more linear) luminosities, as well as a correlation between X-ray and
infrared luminosity, suggesting a link with on-going star formation or
starburst activity.
Optically selected samples of normal galaxies observed with ROSAT
generally confirmed the above picture.
Studying the
X-ray emission of spiral galaxies and analyzing its compact
and diffuse components, \citet*{Read1997} have shown that its origins are
hot diffuse continuum, supernova remnants and high-mass X-ray binaries
(HMXRBs), phenomena which are
related to star formation activity, and that their relative strength is
also related with the galaxy's activity \citep{Read2001}.

The above studies are restricted to optically selected samples of normal
galaxies in the nearby universe. The advent of the second-generation X-ray
telescopes (Chandra and XMM) has allowed for the first time the compilation
of X-ray selected normal galaxy samples. Their selection is
usually based on low X-ray to optical flux ratios
\citep[see][]{Hornschemeier2003} and optical spectral properties
\citep[e.g.][]{Norman2004,Bauer2004}. Searches for normal galaxies in
broad \citep*[e.g.][]{Georgakakis2003,Georgakakis2004,Georgakakis2006,Hornschemeier2005,Tajer2005,Tzanavaris2006}
and deep \citep[e.g.][]{Alexander2002,Hornschemeier2003,Norman2004,Georgakakis2007}
surveys have provided a large number of X-ray selected galaxies.
Using these samples, the luminosity function of normal galaxies has been
established, both locally \citep*[e.g.][]{Georgantopoulos2005,Georgakakis2006}
and at higher redshifts \citep[e.g.][]{Norman2004,Ptak2007,Tzanavaris2008}.

One of the most interesting properties of the X-ray emission of normal
galaxies is its connection to star formation. New results have confirmed
earlier findings that X-rays can act as a star formation indicator in
late-type systems. The tight correlation of the X-ray luminosity with other
star formation indicators, such as infrared and radio luminosities
\citep*{Shapley2001,Ranalli2003} support this
hypothesis. The X-ray emission of late-type galaxies comes from both their
extended structures (spiral arms and the Galactic centre) and compact sources.
Emission from the spiral arms is tightly connected with
star formation \citep{Tyler2004}, while the connection to the Galactic
centre is less well defined. The compact sources on the other hand are mostly
HMXRBs \citep{Colbert2004} which are directly connected to star formation
as end products of massive rapidly evolving stars \citep{Persic2004,Persic2007}.
\citet*{Grimm2003} and \citet*{Gilfanov2004} reproduce the observational
characteristics of X-ray normal galaxies as a combination of discrete point
sources and predict a steep ($L_{\rm x}\propto{\rm SFR}^\beta$ with $\beta>1$)
$L_{\rm x}-{\rm SFR}$
relation when the star formation rate is low
($\lesssim4.5{\rm M}_{\odot}\,{\rm yr}^{-1}$)
and a linear ($\beta=1$) relation thereafter.

In this paper we expand the galaxy samples of \citet{Georgakakis2006} and
explore the relation between the X-ray and H$\alpha$ luminosities for emission
line systems. This provides an insight into the $L_{\rm x}-{\rm SFR}$ relation
\citep[see also][]{Hornschemeier2005}.

\section{Sample Selection}

The first \emph{XMM Newton} Serendipitous Source Catalogue
\citep[1XMM, ][]{Watson2003} is a
compilation of source detections drawn from 585 XMM-Newton EPIC observations
made between March 2000 and May 2002, and released in January 2003. It contains
$\sim55000$ sources, of which $\sim33000$ are considered safe detections and
covers an area of $\sim90$\,deg$^2$. From the 1XMM
catalogue we selected only sources which are observed for more than 10\,ks to
achieve a reasonable signal-to-noise ratio in X-rays and required the detection
likelihood to be above $10\,\sigma$ in the soft band (0.5-2.0\,keV). Since we
are aiming for normal galaxies, we limited our search sample to sources with
hardness ratios $<-0.2$ (between the 0.5-2.0 and 2.0-4.5\,keV bands) to avoid
X-ray obscured AGN. This limit is unlikely to reject any normal galaxies
since their hardness ratios are normally $<-0.2$ (see e.g. Table 1 of
\citealt{Tzanavaris2006}).

To identify these X-ray sources with optical counterparts, we used the USNO-B
optical catalogue \citep{Monet2003}. This is an all-sky catalogue of
$\sim 10^9$ optical sources observed during various surveys over the past
50 years. We searched for optical counterparts to the X-ray sources in the
USNO-B
catalogue with a search radius of 6\,arcsec (see Fig. \ref{optical_images}).
We limited our search to the
southern hemisphere ($\delta <0$) in order to be within the observing range
of the Siding Spring Observatory, and to avoid Galactic stars and high
Galactic $N_H$ we excluded
sources that are within a belt of $\pm20$ degrees of Galactic latitude.
We also avoided the areas of the sky close to the Magellanic Clouds, with an
exclusion radius of 3 and 1.5 degrees from the centres of the LMC and the SMC
respectively. This left us with an initial catalogue of $\sim$16000 XMM
sources, of which 4700 are associated with an optical detection.

A useful diagnostic to discriminate between normal galaxies and AGN is the
X-ray to optical flux ratio $(f_{\rm x}/f_{\rm o})$.
\citet{Hornschemeier2003} found that optically bright X-ray faint objects
(OBXF, which they define as having
$\log(f_{\rm x}/f_{\rm o})<-2.3$)
are almost exclusively galaxies with a contamination from stars. Moreover,
AGN are usually confined in the area
$-1<\log(f_{\rm x}/f_{\rm o})<1$
\citep{Stocke1991,Lehmann2001}.
For the purposes of this study we selected as normal galaxy candidates sources
having $\log(f_{\rm x}/f_{\rm o})<-2$
\citep[see also][]{Georgakakis2004,Georgantopoulos2005,Georgakakis2006}.
This is a conservative limit and it might introduce a bias against massive
ellipticals and powerful starbursts \citep*{Tzanavaris2006,Georgakakis2007}.
On the other hand it minimizes contamination from AGN with low X-ray
luminosities having unusually low $f_{\rm x}/f_{\rm o}$
\citep{Lehmann2001}. To calculate the X-ray to optical flux ratio we used:
\[\log\left(\frac{f_{\rm x}}{f_{\rm opt}}\right)=\log f_{\rm x}+\frac{R}{2.5}+5.5\]
\citep[e.g.][]{Hornschemeier2003}, where $f_{\rm x}$ is the (0.5-2.0)\,keV
X-ray flux in erg\,s$^{-1}$\,cm$^{-2}$, and $R$ is the magnitude from the
USNO-B catalogue.

Selecting our sources taking into account the hardness ratio and the X-ray to
optical flux ratio, left us with a catalogue of 713 normal galaxy candidates
contaminated with a large number of Galactic stars, as they have similar
observational characteristics. We used optical morphologies to identify and
distinguish the stars from the galaxies. For that purpose, we used the APM Sky
Catalogues\footnote{\texttt{http://www.ast.cam.ac.uk/\~\,mike/apmcat/}}
\citep{Irwin1994}, which
are a digitized compilation of the Palomar O and E sky survey in the northern
and the UKST $B_{\rm J}$ sky survey in the southern hemisphere. They also
include
a star-galaxy separation parameter based on the extended nature of each source,
which we used to select extragalactic candidates. Visual inspection of the
sources with the DSS2-Red image \citep{Lasker1996} generally agrees with
the star-galaxy separation from the APM; 98\% of sources characterized as stars
with APM are independently characterized as stars when optically inspecting the
images.

The final catalogue consists of 44 normal galaxy candidates, listed in Table
\ref{candidates}. Their DSS2-Red optical images with an indication on the
location of the X-ray source are shown in Fig.\ref{optical_images}.
The X-ray fluxes of the sources are calculated from the 0.5-2\,keV count rates
given in the 1XMM catalogue, assuming a power-law X-ray spectrum with
$\Gamma=1.9$. In Fig. \ref{fxfo} we plot the $R$ magnitude with respect to the
X-ray flux for emission and absorption line galaxies. We do not observe
any difference between early and late-type galaxies in terms of their relative
X-ray to optical fluxes.

\begin{figure*}[h]
\centering
\resizebox{\hsize}{!}{\includegraphics{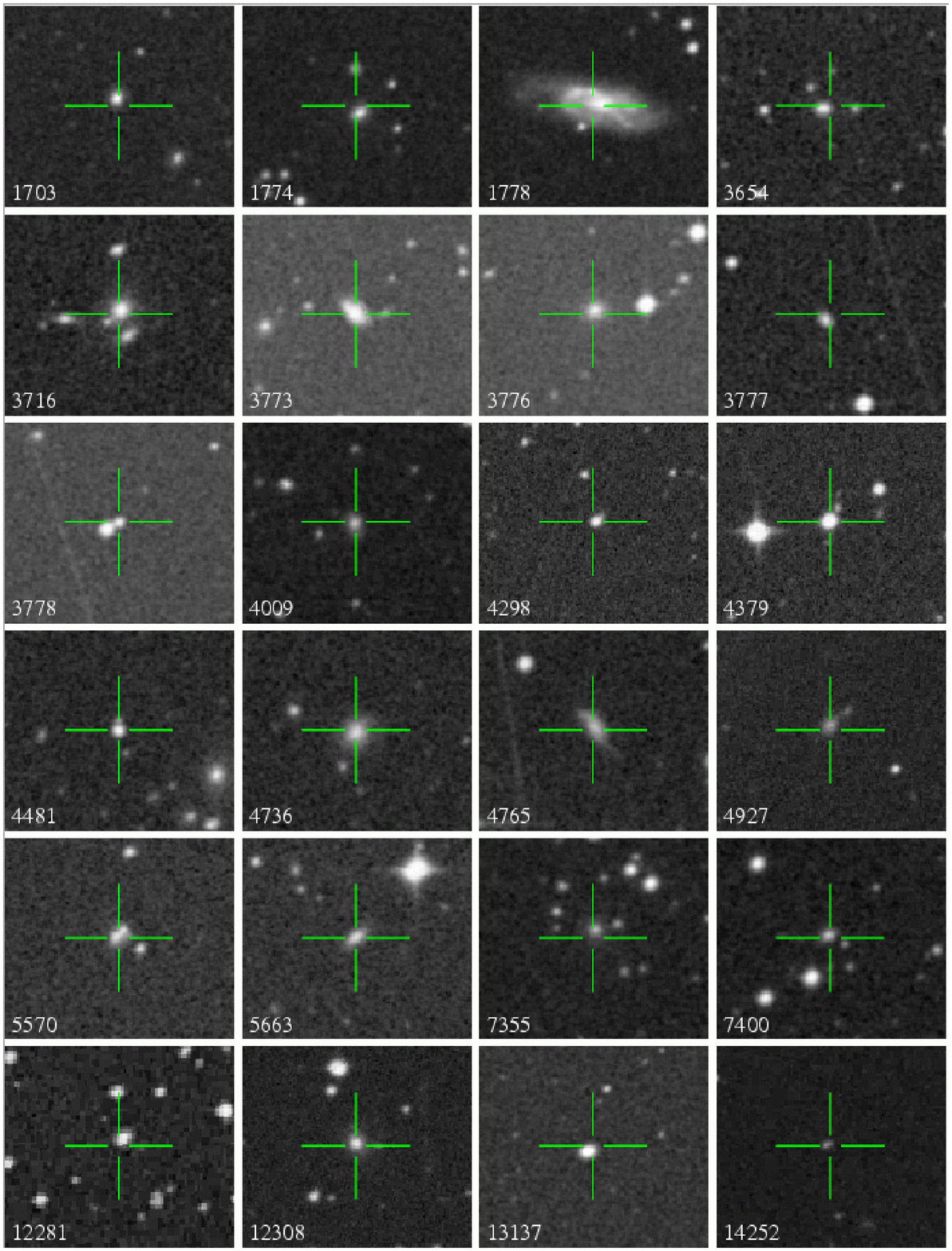}}
\caption{Optical ($R$) images of the normal galaxy candidates, centred on the
         X-ray position. The inner radii of the crosses are 6\,arcsec and the
         outer 30\,arcsec.}
\label{optical_images}
\end{figure*}

\addtocounter{figure}{-1}
\begin{figure*}[h]
\centering
\resizebox{\hsize}{!}{\includegraphics{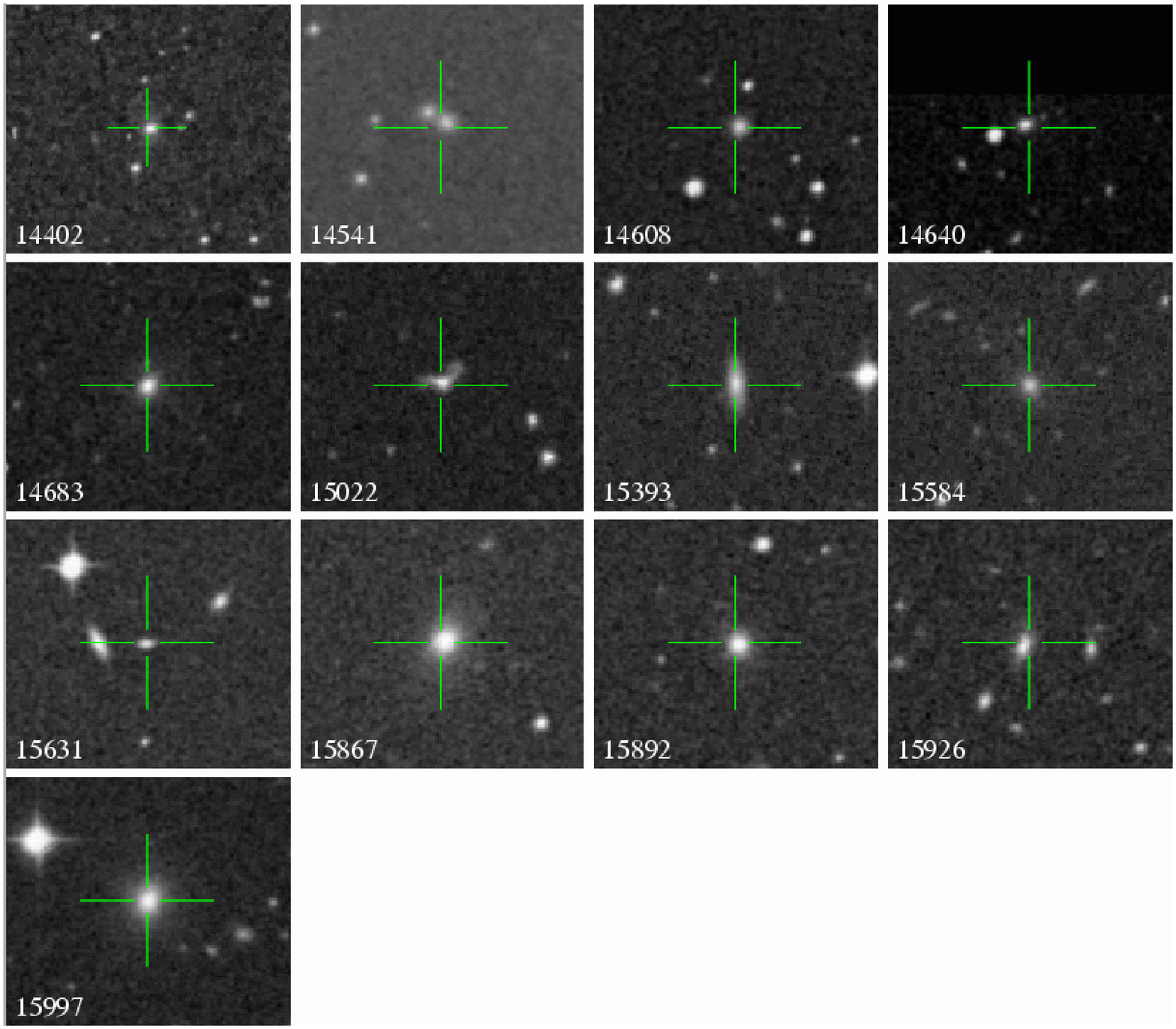}}
\caption{continued\ldots}
\end{figure*}

\begin{figure}[h]
\centering
\resizebox{\hsize}{!}{\includegraphics{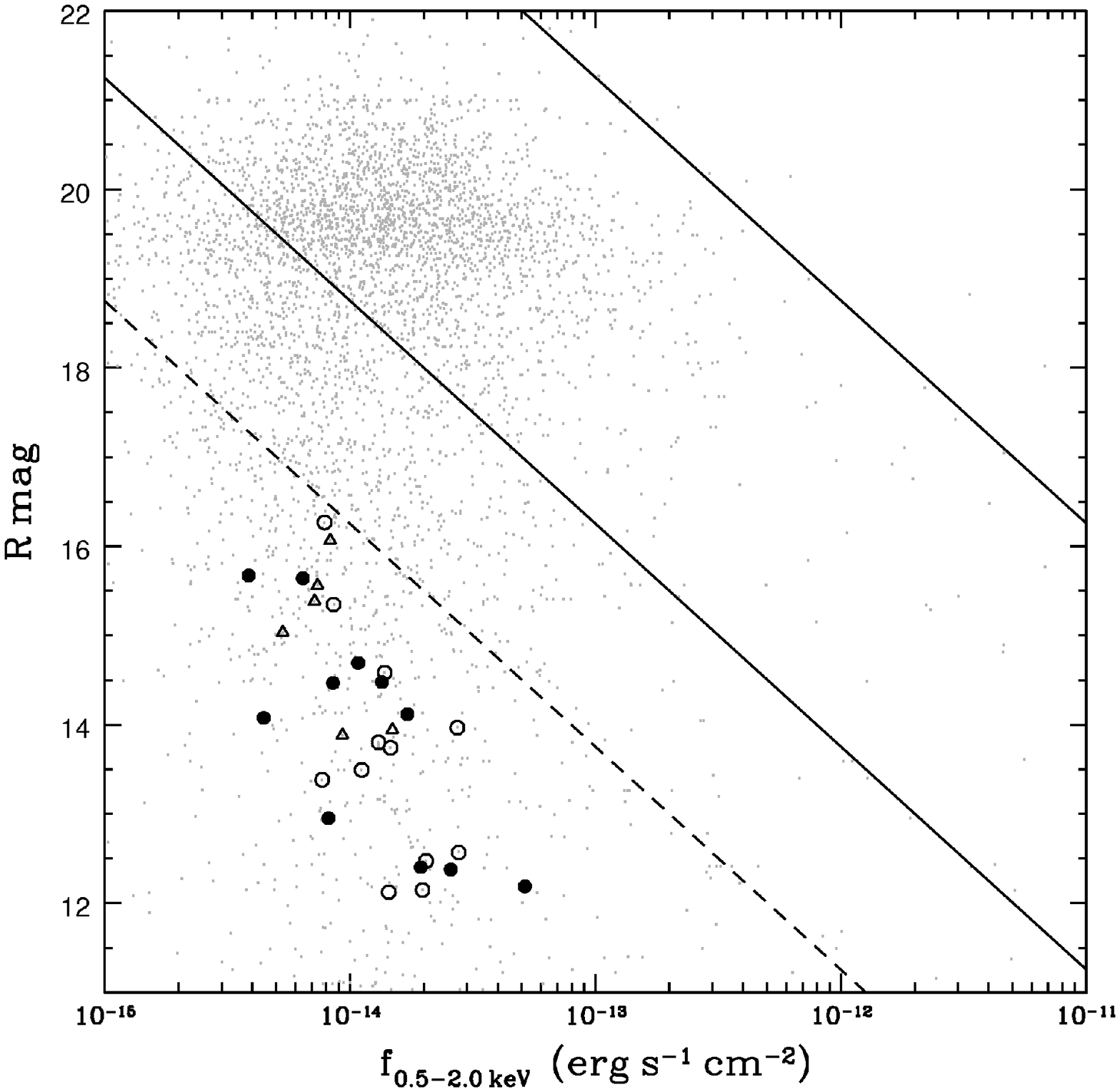}}
\caption{$R$-band magnitude versus X-ray flux for emission and absorption line
         galaxies in our
         sample, marked with filled and open circles respectively. Open
         triangles mark sources with featureless spectra. Solid lines represent
         $f_{\rm x}/f_{\rm o}=1$ and -1, where the bulk of AGN are expected, and
         the dashed line represents $f_{\rm x}/f_{\rm o}=-2$, our limit for
         normal galaxy selection.}
\label{fxfo}
\end{figure}

\section{Observations and Data Reduction}

Long-slit spectral observations were carried out at the 
Australian National University's 2.3\,m telescope
in Siding Spring, Australia, between
2006 October 23-29.
We used the Double Beam Spectrograph (DBS) with
dichroic \#3 and 600 lines/mm gratings. We set
the central wavelength at 5100\,\AA\ (7000\,\AA) in
the blue (red) arm, thus obtaining
continuous coverage over the spectral region
4140\,\AA--7965\,\AA\ with a sampling of $\sim2$\,\AA/pixel
and with a 1$^{\prime\prime}$ slit. As our
observing run was partly affected by
cloudy conditions, we were only able to observe
35 of the full set of 44 normal galaxy candidates.
%
%
Standard calibration exposures were carried out, 
including comparison lamps and spectrophotometric
standard stars for wavelength and flux calibration,
respectively.

\begin{table}
\caption{Observation summary}\label{observations}
\centering
\begin{tabular}{lcccc}
\hline\hline
ID    & XMM name         & obs.date     & exp. time & S/N   \\
      &                  &              & min       &       \\
\hline
1703  & J005818.3-355548 & Oct. 23 2006 & 60        & 32.1  \\
1774  & J005922.8-360933 & Oct. 26 2006 & 45        &  3.6  \\
1778  & J005929.7-361113 & Oct. 28 2006 & 30        & 34.0  \\
3654  & J022416.7-050323 & Oct. 29 2006 & 55        & -     \\
3716  & J022456.2-050801 & Oct. 24 2006 & 40        &  2.5  \\
3773  & J022536.4-050012 & Oct. 24 2006 & 30        & 84.7  \\
3776  & J022537.8-050223 & Oct. 24 2006 & 40        & -     \\
3777  & J022538.2-050806 & Oct. 25 2006 & 45        & 118.0 \\
3778  & J022538.3-050423 & Oct. 29 2006 & 30        & 3.4   \\
4009  & J023613.5-523036 & Oct. 23 2006 & 40        & 131.4 \\
4298  & J030927.5-765223 & Oct. 28 2006 & 40        & 4.5   \\
4379  & J031256.5-765039 & Oct. 29 2006 & 30        & 9.4   \\
4481  & J031723.1-442056 & Oct. 28 2006 & 40        & 4.5   \\
4736  & J031829.8-441140 & Oct. 29 2006 & 45        & 23.7  \\
4765  & J031845.0-441042 & Oct. 23 2006 & 55        & 11.4  \\
4927  & J033831.4-351421 & Oct. 25 2006 & 40        & 6.8   \\
5570  & J043306.5-610760 & Oct. 24 2006 & 55        & 3.3   \\
5663  & J043333.5-612427 & Oct. 24 2006 & 40        & 2.3   \\
7355  & J055940.7-503218 & Oct. 24 2006 & 45        & 2.7   \\
7400  & J060014.9-502230 & Oct. 24 2006 & 30        & -     \\
12281 & J201329.7-414737 & Oct. 27 2006 & 45        & 79.2  \\
12308 & J201345.0-563713 & Oct. 23 2006 & 30        & 2.0$^1$   \\
13137 & J213758.7-143611 & Oct. 23 2006 & 30        & 61.6  \\
14402 & J221726.0-082531 & Oct. 26 2006 & 30        & 2.6   \\
14541 & J222110.0-244749 & Oct. 28 2006 & 40        & -     \\
14608 & J222804.4-051751 & Oct. 29 2006 & 45        & -     \\
15022 & J225149.3-175225 & Oct. 23 2006 & 30        & 153.7 \\
15393 & J231421.6-424559 & Oct. 24 2006 & 40        & 4.0   \\
15584 & J231851.8-423114 & Oct. 25 2006 & 30        & 2.9   \\
15631 & J232454.9-120459 & Oct. 28 2006 & 45        & 1.9   \\
15867 & J235340.6-102420 & Oct. 23 2006 & 30        & 3.2   \\
15892 & J235405.7-101829 & Oct. 27 2006 & 45        & 1.3   \\
15926 & J235418.1-102013 & Oct. 28 2006 & 40        & 14.3  \\
15997 & J235629.1-343743 & Oct. 25 2006 & 30        & 2.0$^1$   \\
\hline
\end{tabular}
\begin{enumerate}
\item{S/N of the Mg I b2 line}
\end{enumerate}
\end{table}

Standard IRAF\footnote{IRAF is distributed by the 
National Optical Astronomy Observatories,
which is operated by the Association of Universities for Research in
Astronomy, Inc. (AURA) under cooperative agreement with the National
Science Foundation.}
routines were used for data reduction. 
%
All frames were overscan and bias-subtracted.
Observations at the same setting were combined
to increase the signal-to-noise ratio (S/N) and to
remove cosmic rays. 
Subsequently, the spectra were
flat-fielded, sky-subtracted, wavelength- and flux-calibrated.
Details of the observations can be found in Tab. \ref{observations}.
The signal-to-noise ratio is the ratio of the most prominent
feature of each spectrum, which for emission-line systems is
the H$\alpha$ line and for absorption-line systems the
Na D doublet. 

\section{Optical Spectral Properties}

\begin{figure*}[h]
\centering
\resizebox{\hsize}{!}{\includegraphics{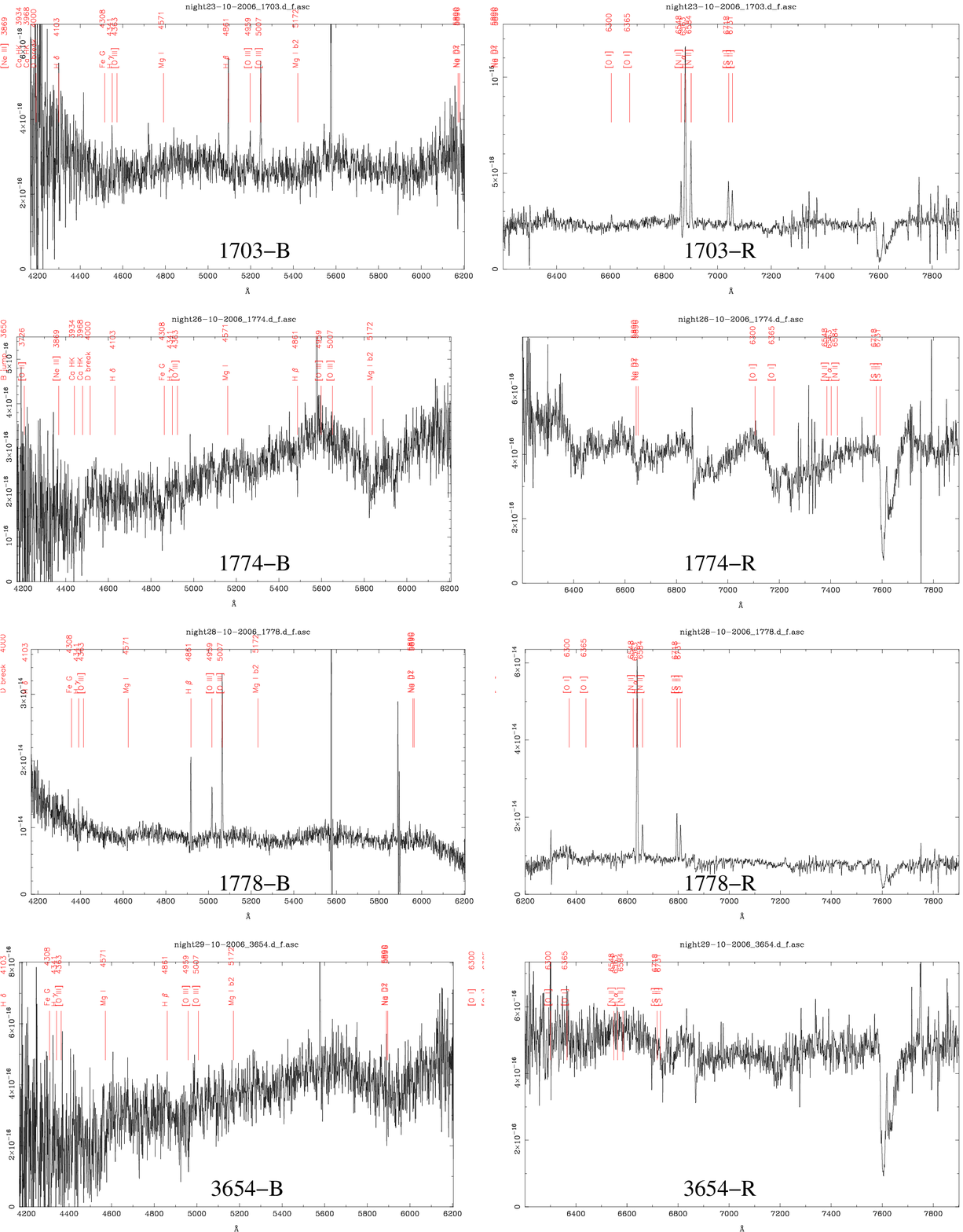}}
\caption{Optical spectra of the normal galaxy candidates.}
\label{spectra}
\end{figure*}

\addtocounter{figure}{-1}
\begin{figure*}[h]
\centering
\resizebox{\hsize}{!}{\includegraphics{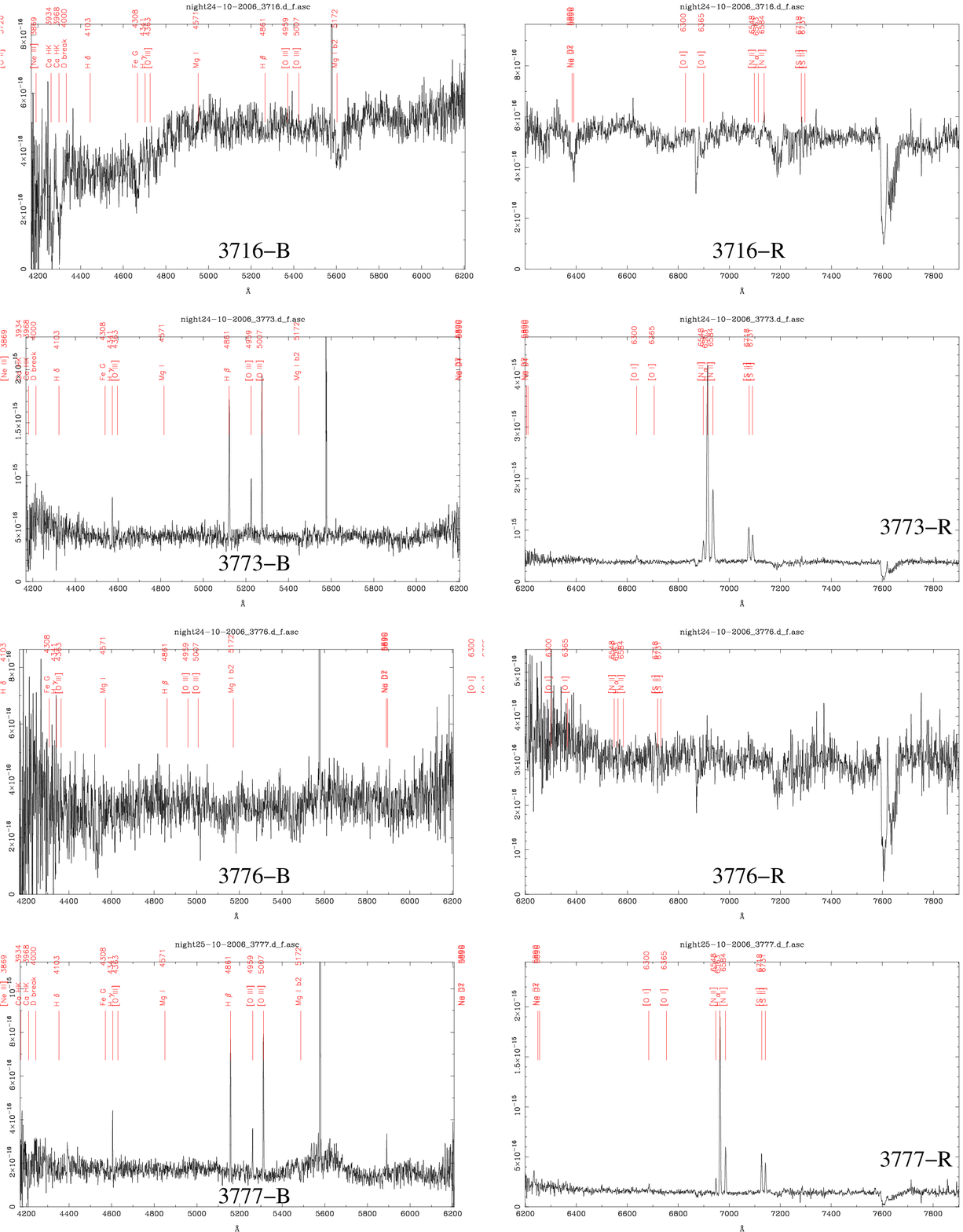}}
\caption{Continued\ldots}
\end{figure*}

\addtocounter{figure}{-1}
\begin{figure*}[h]
\centering
\resizebox{\hsize}{!}{\includegraphics{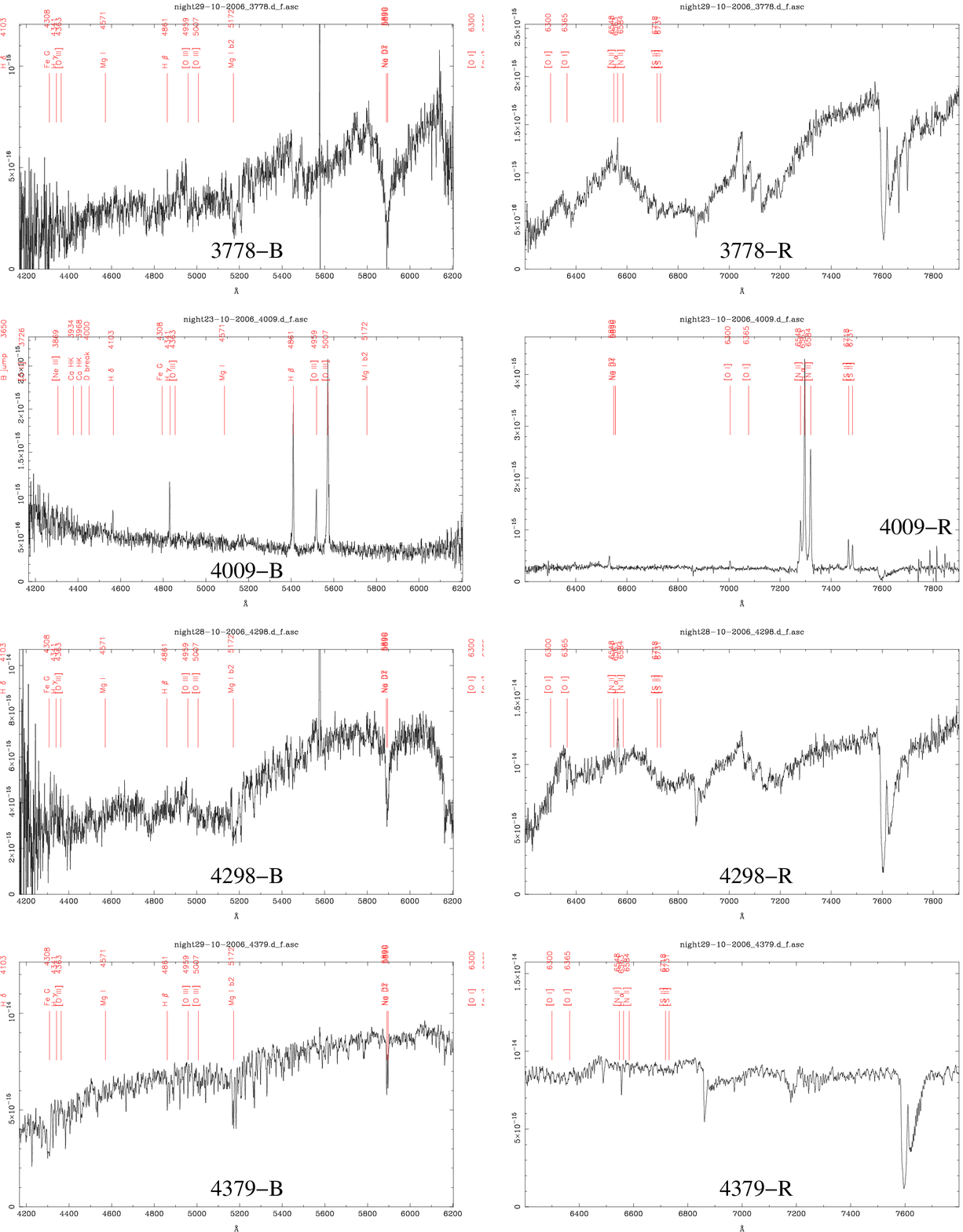}}
\caption{Continued\ldots}
\end{figure*}

\addtocounter{figure}{-1}
\begin{figure*}[h]
\centering
\resizebox{\hsize}{!}{\includegraphics{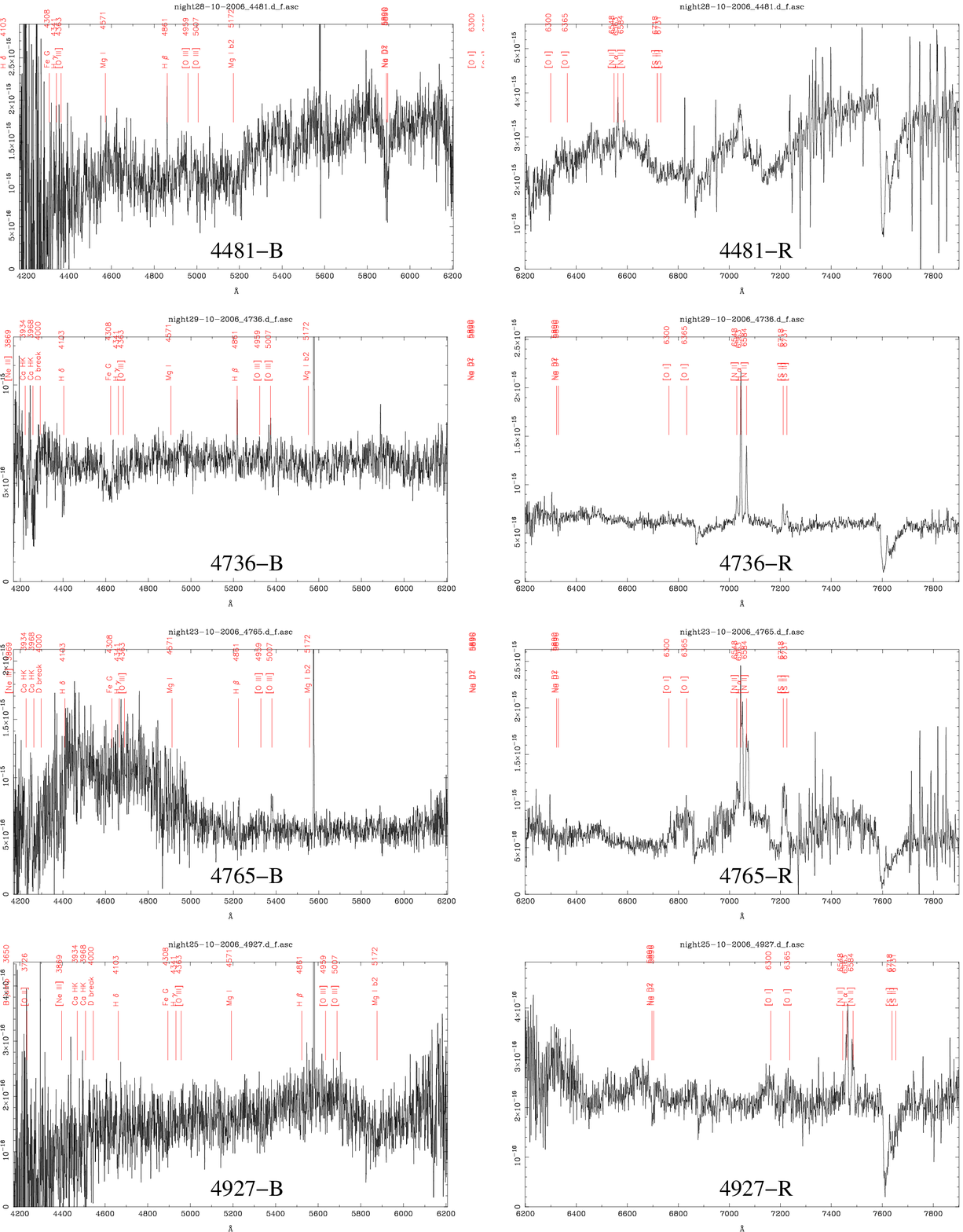}}
\caption{Continued\ldots}
\end{figure*}

\addtocounter{figure}{-1}
\begin{figure*}[h]
\centering
\resizebox{\hsize}{!}{\includegraphics{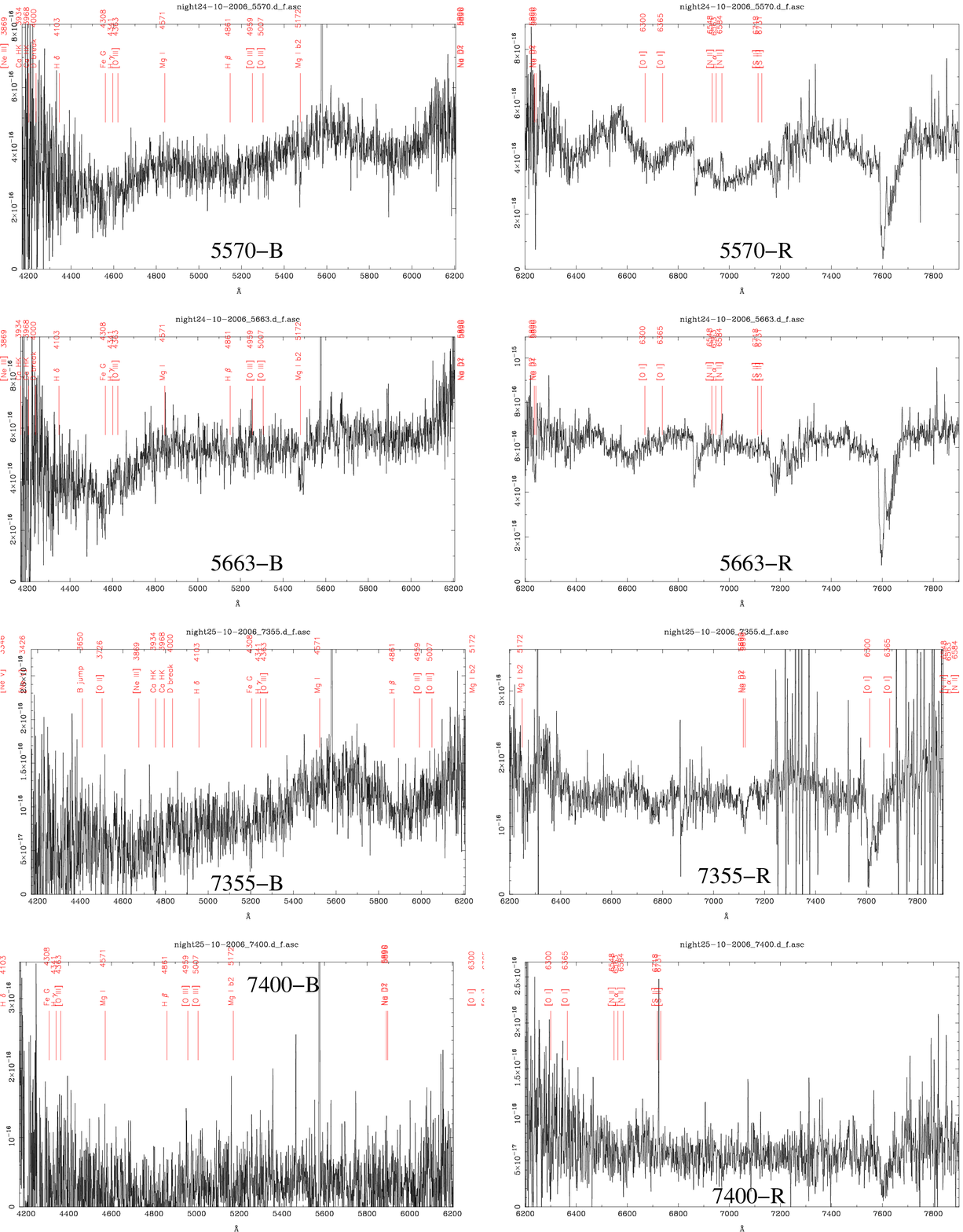}}
\caption{Continued\ldots}
\end{figure*}

\addtocounter{figure}{-1}
\begin{figure*}[h]
\centering
\resizebox{\hsize}{!}{\includegraphics{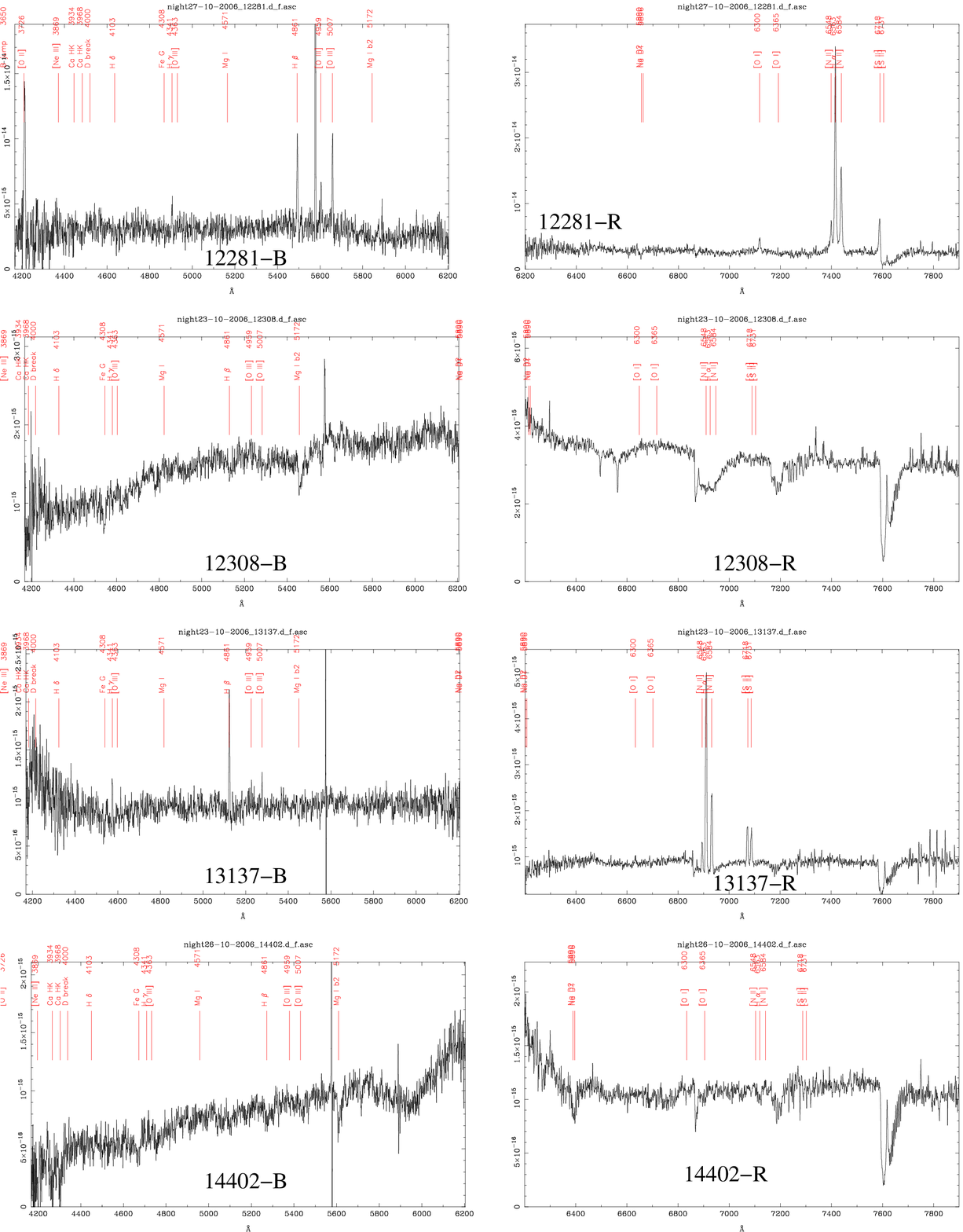}}
\caption{Continued\ldots}
\end{figure*}

\addtocounter{figure}{-1}
\begin{figure*}[h]
\centering
\resizebox{\hsize}{!}{\includegraphics{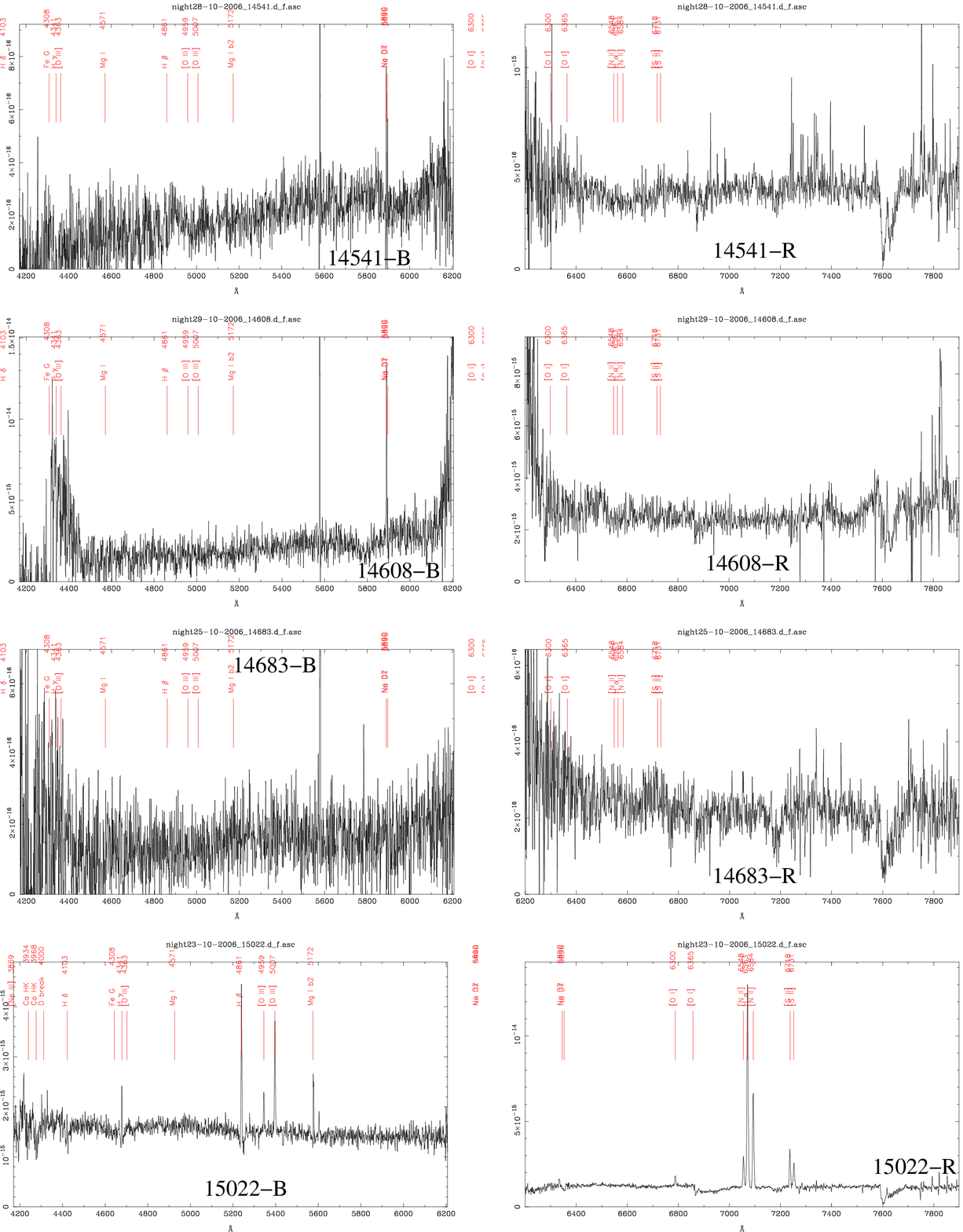}}
\caption{Continued\ldots}
\end{figure*}

\addtocounter{figure}{-1}
\begin{figure*}[h]
\centering
\resizebox{\hsize}{!}{\includegraphics{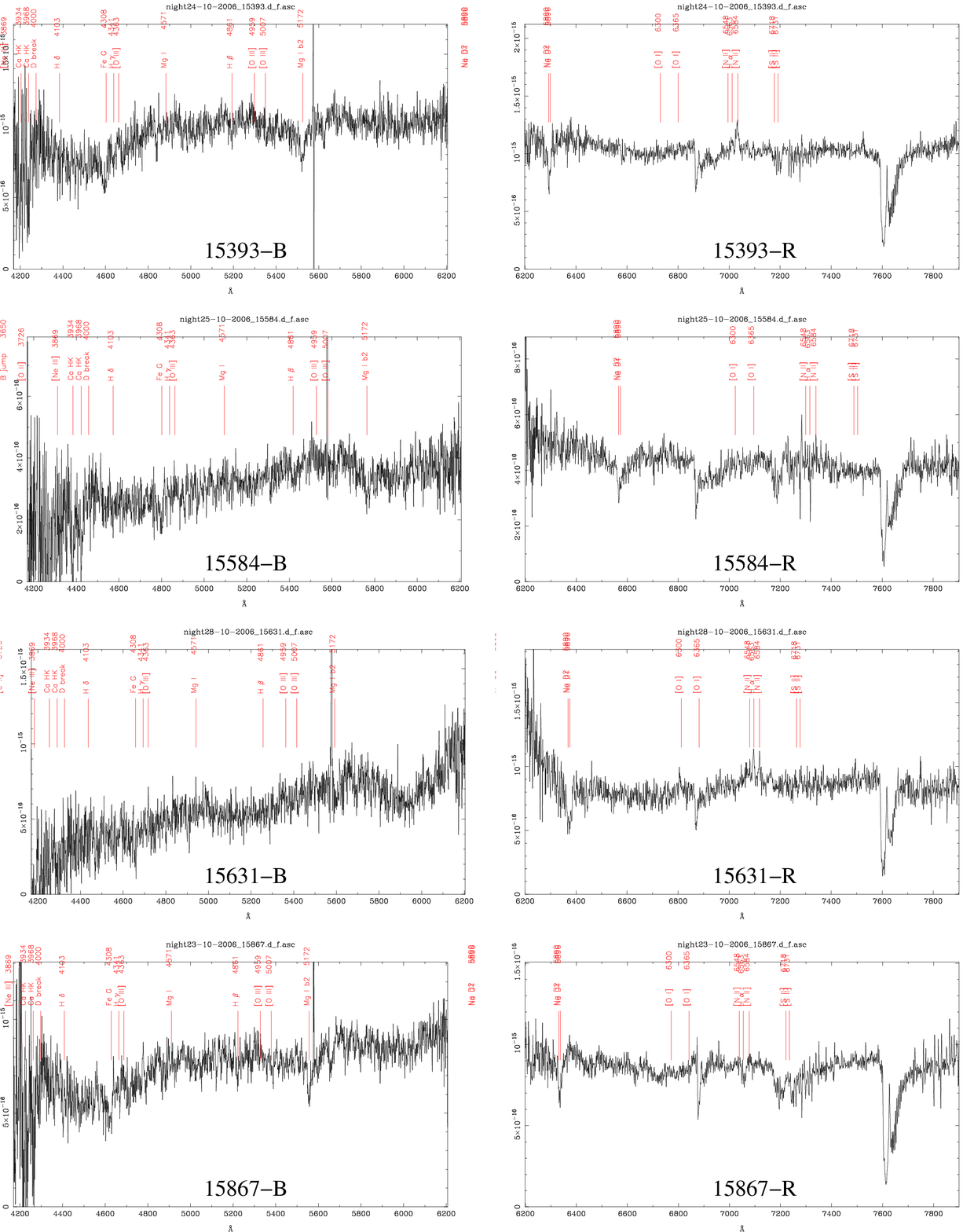}}
\caption{Continued\ldots}
\end{figure*}

\addtocounter{figure}{-1}
\begin{figure*}[h]
\centering
\resizebox{\hsize}{!}{\includegraphics{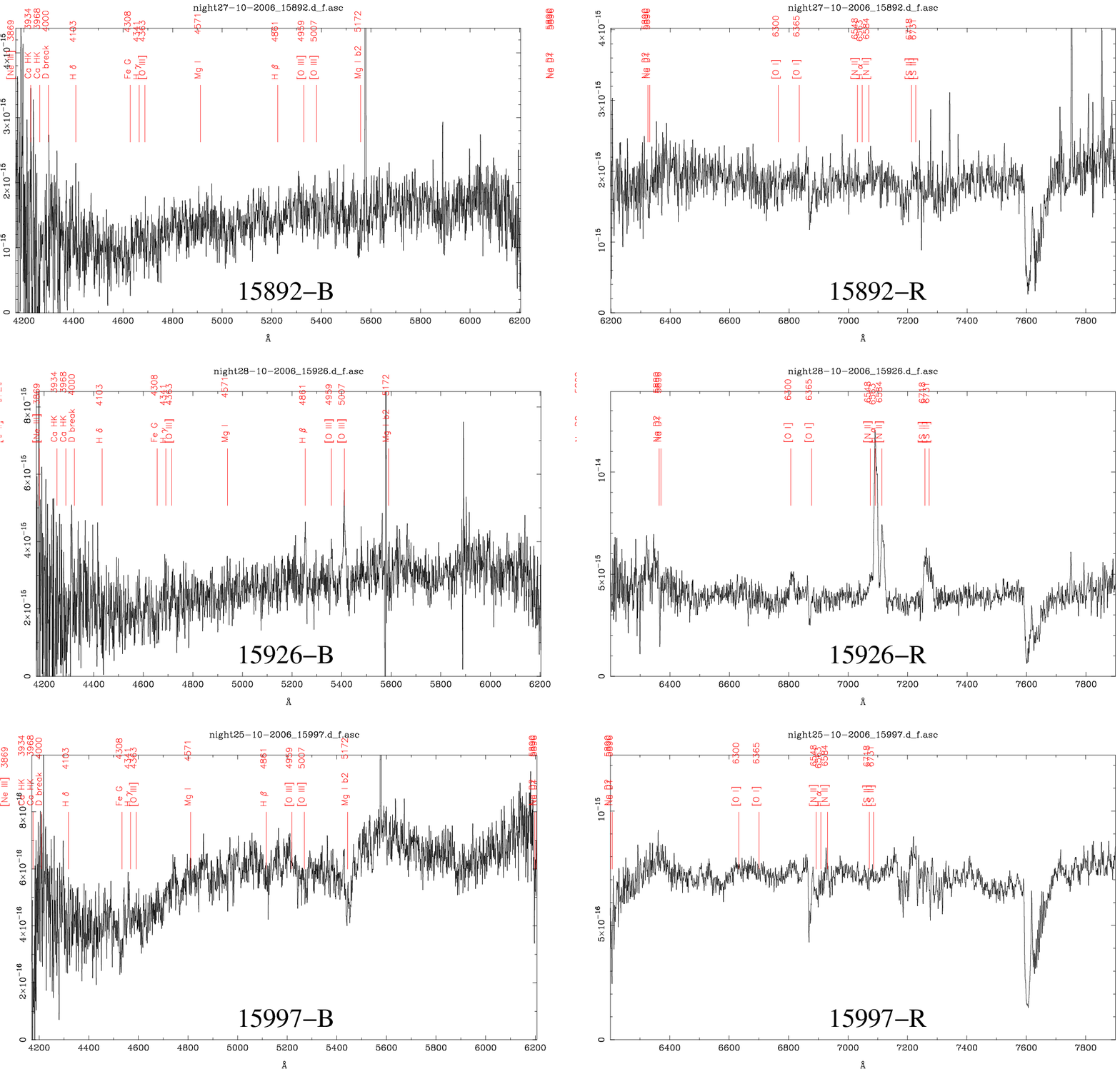}}
\caption{Continued\ldots}
\end{figure*}

We have obtained spectra for 35 of the normal galaxy candidates, which are
shown in Fig.\ref{spectra}. For 29 sources we were able to identify emission
and absorption lines, the most prominent
of which are: H$\alpha$, H$\beta$, H$\gamma$, [S\,II](6718\,\AA),
[S\,II](6731\,\AA), [O\,I](6300\,\AA), [O\,III](4959\,\AA), [O\,III](5007\,\AA),
[N\,II](6548\,\AA), [N\,II](6584\,\AA)
in emission, and Na\,D1(5898\,\AA), Na\,D2(5892\,\AA), Mg\,I\,b2(5172\,\AA),
Fe\,G(4308\,\AA), Ca\,HK(3934 \& 3968\,\AA) in absorption. 11 galaxies are
characterized as emission line systems, 12 absorption line systems, while 2
sources show both emission and absorption lines and are characterized as
composite and 4 sources have stellar-like spectra. The redshifts of the sources
with identified lines were calculated from the rest-frame frequencies by fitting
gaussians to the line profiles. They agree with published values.
7 of the redshifts we obtained are not published in the literature.

In cases of systems with emission lines, we measured their intensities and
used the line ratios to trace signs of AGN. For this purpose, we use the line
ratio diagnostics of \citet*{Ho1997}, which are based on the ratios of the
[O\,III](5007\,\AA), [O\,I](6300\,\AA), [N\,II](6584\,\AA), [S\,II](6718\,\AA)
and [S\,II](6731\,\AA) lines with the Balmer lines (H$\alpha$ and H$\beta$).
These line combinations ([O\,III]/H$\beta$ - [O\,I]/H$\alpha$ - [N\,II]/H$\alpha$ -
[S\,II]/H$\alpha$) are selected to have small wavelength separation and
therefore be insensitive to reddening \citep[see also][]{Veilleux1987}.
Without including any reddening correction,
we find that the emission line ratios are compatible with starburst galaxies
for all the emission line systems. For systems having composite spectra, the
emission line ratios are compatible with Seyfert nuclei. These are removed from
our sample of normal galaxies.

Sources which show no prominent features in their spectra are marked as
``featureless'' in Table \ref{candidates}. A featureless optical spectrum
can indicate a BL Lac object \citep[e.g.][]{Londish2002,Sbarufatti2006}.
The broadband spectra of BL Lacs show a variety of X-ray to
optical to radio flux ratios \citep{Bondi2001} and there are cases where BL
Lacs are bright in optical and faint in X-rays \citep{Troitsky2008}. The host
galaxies
of BL Lacs are in most cases ellipticals or bulge dominated systems
\citep{Falomo1996} and optical images are consistent with this picture.
We therefore
cannot rule out that at least some of the sources with featureless spectra
harbor AGN and we remove them from our final sample of normal galaxies.

\section{Starburst diagnostics}

\begin{figure}[h]
\centering
\resizebox{\hsize}{!}{\includegraphics{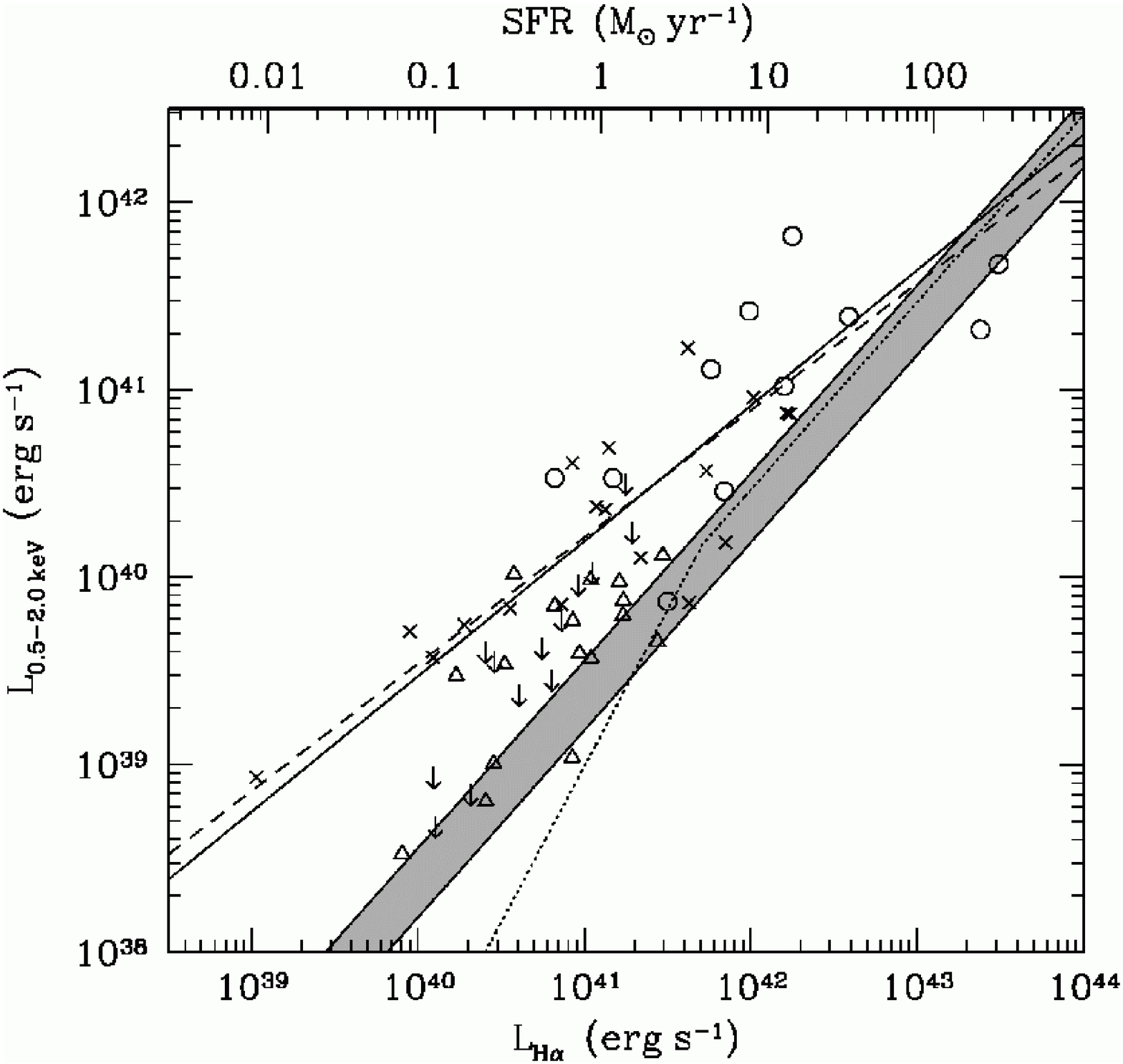}}
\caption{X-ray luminosity (in the 0.5-2\,keV band) plotted against the
         luminosity of the H$\alpha$ line for late-type galaxies. With circles
         are plotted the data-points of this study and with crosses the
         data-points of \citet{Georgakakis2006}. Triangles and arrows represent
         the data-points (and respective upper limits) of the combined sample of
         local spirals from \citet{Shapley2001} and \citet{Trinchieri1989}.
         The solid and dashed lines
         are the fits to the combined dataset using the bisector line and the
         orthogonal regression respectively. The shaded region represents the
         relations of \citet{Ranalli2003} and \citet{Persic2007} and the dotted
         line the relation of \citet{Grimm2003}. The SFR scale is also shown
         for comparison on the top axis.
         }
\label{lxlha}
\end{figure}

A good diagnostic of the star formation rate is the luminosity of the H$\alpha$
line, which is found to scale linearly with the SFR for nearby galaxies and
also at higher redshifts \citep*{Kennicutt1994,Kewley2002,RosaGonzalez2002}.
The X-ray versus the H$\alpha$ luminosity diagram is shown in Fig. \ref{lxlha}.
We corrected the derived H$\alpha$ luminosity for dust extinction using
the reddening curves of \citet{Savage1979} and the flux ratio of the
H$\alpha$ and H$\beta$ lines, assuming an intrinsic value of 2.76
\citep{Brocklehurst1971}. Our data points are plotted as open circles in Fig.
\ref{lxlha}, while crosses represent the combined sample of
the 1XMM survey of \citet{Georgakakis2006} and the NHS survey 
\citep{Georgakakis2004}.

Because our dataset has errors in both directions we choose a fitting method
that deals with the two axes in a symmetric way. As such we choose the
bisector line and the orthogonal regression method \citep{Isobe1990}, which
yield $\beta=0.72\pm 0.04$ and $\beta=0.68\pm 0.05$ respectively when
fitting  $L_{\rm x}=\alpha L_{\rm H\alpha}^{\beta}$. These are the solid and
dashed lines in Fig. \ref{lxlha}. 
These values lie within the uncertainty ranges of the measurements made by
\citet{Griffiths1990} ($0.70\pm0.12$), \citet{Zezas2000} ($0.62\pm0.11$), and
\citet{Georgakakis2006} ($0.69\pm0.06$)
and they are not linear. If we
attribute the X-ray emission to a number of HMXRBs, the relation is expected
to be linear \citep{Persic2004,Persic2007}. Indeed, a linear relation is found
by \citet*{David1992} and \citet{Ranalli2003} for both soft and hard X-rays,
using the far-infrared and radio (1.4\,GHz) as star formation indicators.
Non-linearity
is expected in cases of very low or very high star
formation rates, where, as a result of small number statistics and hypothetical
intermediate mass black holes respectively, it is expected to be even
steeper with $\beta>1$ \citep{Grimm2003,Gilfanov2004}. If we use the relation:
SFR($M_\odot$\,yr$^{-1}$)=$7.9\times 10^{-42}L_{\rm H\alpha}$(erg\,s$^{-1}$)
\citep{Kennicutt1994}, we are expecting the low-SFR cut at
$L_{\rm H\alpha}\sim10^{42}$\,erg\,s$^{-1}$, which is within the range we are
sampling (see dotted line in Fig. \ref{lxlha}). It is therefore puzzling that
the $L_{\rm x}-L_{\rm SFR}$ rate appears flatter than linear. 

Such a behavior could be explained by the existence of other sources
of X-ray emission not connected with star formation activity. As such we could
consider LMXRBs, older supernova
remnants or globular clusters \citep*[see e.g.][]{Vogler1997} and hot diffuse
gas toward the Galactic centre \citep{Tyler2004}. If not linked with star
formation, their relative contribution to the total X-ray luminosity would be
stronger in galaxies with lower
star formation rates, thus flattening the $L_{\rm x}-L_{\rm SFR}$ relation.

The combined X-ray luminosity of LMXRBs is associated with the stellar
content of the galaxy \citep{Gilfanov2004b}, which can be examined from its
luminosity in the K-band. For that purpose we search for counterparts of the
galaxies in our sample in the 2MASS survey catalogues \citep{Skrutskie2006}.
We then calculate the K-band luminosities and use the mass-to-light ratio
of \citet{Bell2001} to derive the stellar mass of each source. This is
translated into the respective luminosity of the LMXRB content using the
relation of \citet{Gilfanov2004}; the results are presented in Tab.
\ref{candidates}. We find that the combined X-ray luminosity of LMXRBs is two
orders of magnitude fainter than the total X-ray luminosity and thus not
enough to have an effect on the $L_{\rm x}-L_{\rm SFR}$ correlation.

\begin{figure}[h]
\centering
\resizebox{\hsize}{!}{\includegraphics{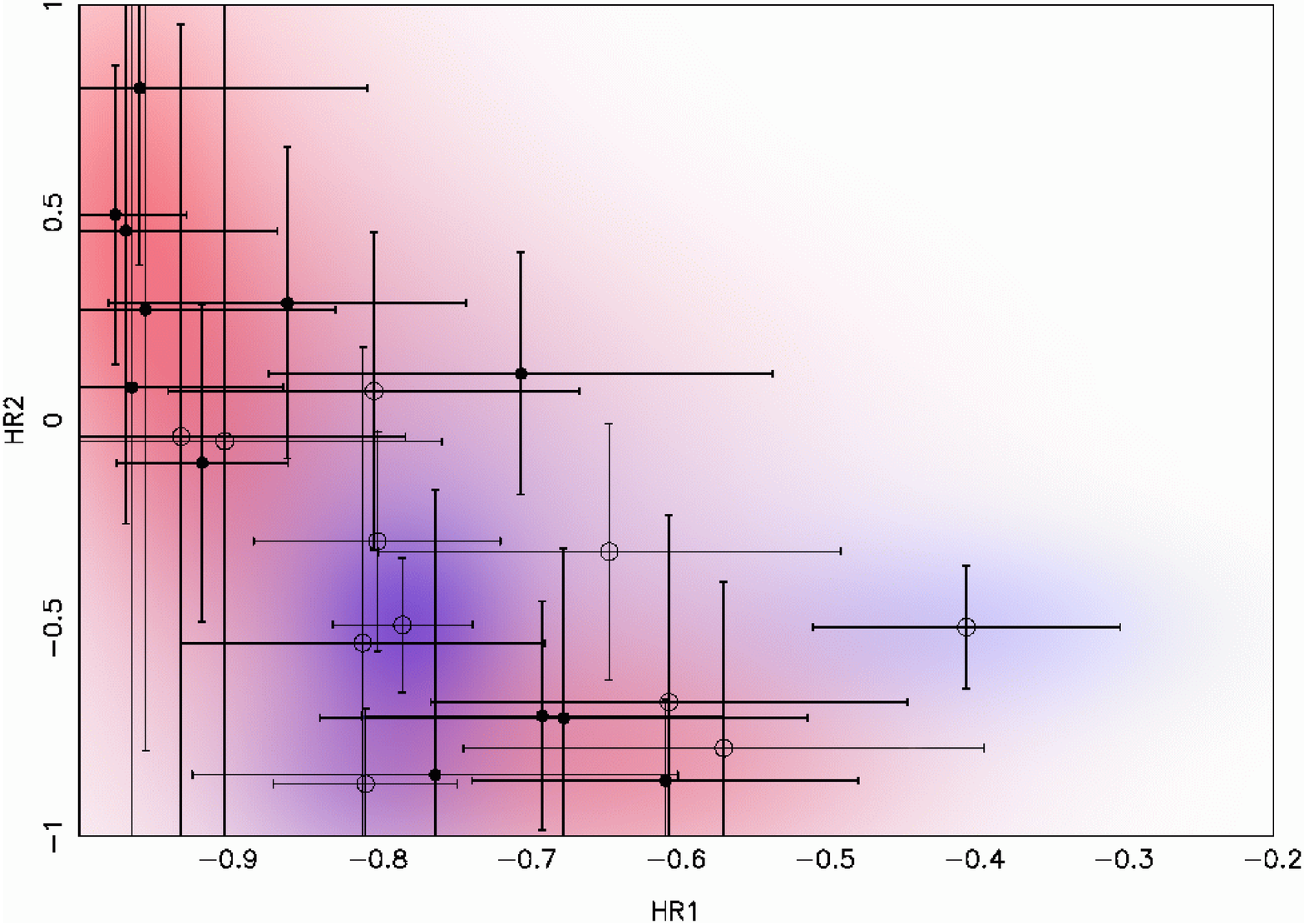}}
\caption{X-ray colour-colour diagram. HR1 and HR2 are the hardness ratios
         between the $(0.5-2.0)$\,keV and $(2.0-4.5)$\,keV, and
         $(2.0-4.5)$\,keV and $(4.5-7.5)$\,keV bands respectively.
         Open circles represent late-type and filled circles early-type
         galaxies.
         Blue and red coloured gaussians are plotted at each late and early
         type point respectively, with their standard deviations equal to the
         uncertainty of the hardness ratio of each point.}
\label{hardness}
\end{figure}

In Fig. \ref{hardness} we plot two different hardness ratios for emission and
absorption like systems. HR1 refers to the hardness ratio between the
$(0.5-2.0)$\,keV and $(2.0-4.5)$\,keV bands, whereas HR2 refers to the
$(2.0-4.5)$\,keV
and $(4.5-7.5)$\,keV bands. As different sources have been observed with
\emph{XMM-Newton} in different modes with different CCDs and different optical
light blocking filters, to have consistent hardness ratios, we
transformed all count rates to the PN CCD with thin filter mode, correcting
for the respective
galactic absorption of each source. Late-type galaxies are represented by blue
colour and early-type by red. The errors in the count rates of each source
have been taken into account to create a 2 dimensional gaussian to be used in
creating the colour representation of Fig. \ref{hardness}. 
All sources by selection have ${\rm HR}1<-0.2$. We can see that the bulk of
the late-type galaxies appear soft in both hardness ratios, which is not a
typical behavior of HMXRBs, therefore a contribution of another
X-ray source should be considered.

In studies of nearby galaxies, where the X-ray emission from point sources
can be resolved out, the diffuse emission from hot gas is generally spatially
coincident with star formation regions
\citep[e.g.][]{Doane2004,Warwick2007,Owen2009}, consistent with its heating
mechanism being supernova explosions and the winds of very massive stars,
possibly with a small contribution from the bulge \citep{Tyler2004}. The
origin of it in star forming related properties would make its contribution
scale linearly with the star formation rate and not affect the slope of the
$L_{\rm x}-L_{\rm SFR}$ relation. \citet{Persic2004} find that considering
only the X-ray contribution of the point sources in their sample of star
forming galaxies, that affects only the scatter of the $L_{\rm x}-L_{\rm SFR}$
relation (making it smaller) and not the slope which is in both cases linear.

In order to test the non-linearity of the $L_{\rm x}-L_{\rm SFR}$ relation, we
attempt to reproduce it using other proxies of star formation rate than the
H$\alpha$ luminosity.
The far infrared luminosity is a good example and we search
for FIR counterparts of the late-type galaxies in the combined sample of
this study and that of \citet{Georgakakis2006} in the IRAS faint source
catalogue \citep{Moshir1990}. We detect 8 sources and
calculate their FIR luminosities using the formula of \citet*{Helou1985}.
We find tentative evidence that the $L_{\rm x}-L_{\rm FIR}$ is flatter than
linear ($b=0.81\pm0.09$ and $b=0.76\pm0.11$ using again the bisector and the
orthogonal regression lines). However the number of data points is too small to
allow us to extract conclusive results; moreover earlier studies of the
$L_{\rm x}-L_{\rm FIR}$ relation using larger samples \citep[e.g.][]{Fabbiano2002}
are in agreement with a linear relation.

We must caution here two effects that might bias our data. Our sample and also
these of \citet{Georgakakis2004} and \citet{Georgakakis2006} are selected
using an $f_{\rm x}/f_{\rm o}$ cut-off and are thus flux limited in the X-rays.
A flux limited sample could introduce a bias in the $L_{\rm x}-L_{\rm SFR}$ relation
flattening it. Also, the star formation rate is proxied by the luminosity of
the H$\alpha$ line. The H$\alpha$ flux is measured through slit spectroscopy
and this technique allows only a fraction of the light of the source to be
detected. The slit width is 1$^{\prime\prime}$, which corresponds to 3.4\,kpc
at a redshift of 0.21 and to 0.23\,kpc at a redshift of 0.0115. This is
significantly smaller than the typical diameter of a spiral galaxy.
Moreover, this effect is more severe for galaxies at lower redshift, which are
the less luminous in flux limited surveys. This introduces a bias in the
$L_{\rm x}-L_{\rm SFR}$ relation and more specifically it flattens it. If we
check the normalization of the $L_{\rm x}-L_{\rm SFR}$ we derive with respect to
\citet{Ranalli2003} and \citep{Persic2007} (shaded region in Fig.
\ref{lxlha}, see also \citealt{Hornschemeier2005} and references therein), we
find an agreement for luminous sources (at higher redshifts) and a deviation
of fainter sources (at lower redshifts). Interestingly, the normalization
in the case of \citet{Hornschemeier2005} where there is a correction applied
to the H$\alpha$ fluxes to compensate for the aperture effect seems to be in
good agreement with  \citet{Ranalli2003} and \citet{Persic2007}.

To further investigate the aperture effect to our data, we compile a sample of
(nearby) normal galaxies with measured X-ray and integrated H$\alpha$ luminosities.
We use the EINSTEIN spiral galaxy sample of \citet{Shapley2001} and combine it
with the sample of \citet*{Trinchieri1989}, which uses integrated H$\alpha$
measured from \citet{Kennicutt1983}. We find 30 common sources between the two
samples, not including sources from the \citet{Shapley2001} sample which show
evidence for non-stellar ionizing source. We transform the X-ray luminosities
to the 0.5 - 2.0\,keV band using a power-law profile with $\Gamma=1.9$. We also
adopt the distances from \citet{Shapley2001} and transform the H$\alpha$
luminosities accordingly. The $L_{\rm x}-L_{\rm H\alpha}$ pairs are plotted with
triangles in Fig. \ref{lxlha} (arrows indicate upper $L_{\rm x}$ limits). We can
see that these data-points lie close to the shaded area and are systematically
more H$\alpha$ luminous compared to the data-points of this study, and the bisector
line fit (not taking upper limits into account) has a slope of
$\beta=1.05\pm0.10$. We therefore conclude that it the ``aperture effect'' of
slit spectroscopy plays an important role in the flattening of the
$L_{\rm x}-L_{\rm SFR}$ relation.

\section{Summary and Conclusions}

In this study we select normal galaxies from the XMM-Newton first serendipitous
source catalogue (1XMM). We select soft sources with reliable detections and
compare their X-ray with their optical fluxes. After selecting candidates with
$\log(f_{\rm x}/f_{\rm o})<-2$ we examine their morphologies to remove any
contamination from Galactic stars. This way we are left with a sample of 44
sources (in the southern hemisphere). We observed 35 of them with the ANU's
2.3\,m telescope to derive their optical spectra. Of the 35 sources, 4 are
Galactic stars, 2 are associated with AGN, and 6 have featureless spectra and
are probably associated with BL-Lac objects. The remaining 23 (65.7\%) are
indeed normal galaxies with emission (11) and absorption-line (12) spectra. 

We examined the sub-sample of emission line galaxies (late type) to derive
their star formation properties. For this purpose we combined our sample with
those of \citet{Georgakakis2004} and \citet{Georgakakis2006}. We find
tentative evidence that the X-ray luminosity does not scale linearly with star
formation, approximated by the H$\alpha$ luminosity, this result is however
sensitive to observational and selection biases. The normalization of the
$L_{\rm x}-L_{\rm SFR}$ relation agrees with previous studies at the luminous
end, indicating underestimation of the H$\alpha$ luminosities for fainter
sources.

\begin{acknowledgements}
ER wishes to thank the European Social Fund (ESF), Operational Program for
Educational and Vocational Training II (EPEAEK II), and particularly
the Program PYTHAGORAS II, for funding part of this work.
\end{acknowledgements}

\begin{table*}
\caption{X-ray, optical and near-infrared properties of the initial candidates.}\label{candidates}
\centering
\begin{tabular}{lcccccccccc}
\hline\hline
ID & XMM name & $z$ & $f_{{\rm x}}$ & $\log L_{\rm x}$ & $R$ & $\log\frac{f_{\rm x}}{f_{\rm o}}$ & $L_{\rm H\alpha}$ & $K$ & $L_{\rm K}$ & type \\
   &          &     & erg\,cm$^{-2}$\,s$^{-1}$ & erg\,s$^{-1}$ & mag &                                     & erg\,s$^{-1}$    & mag & erg\,s$^{-1}$     \\
\hline
1703  & J005818.3-355548 &  $0.0489\pm0.0008$  & $6.421\times10^{-15}$ & 40.53 & 15.64 & -2.436 & 40.82 & 14.585 & 42.84 & emission     \\
1774  & J005922.8-360933 &  $0.1271\pm0.0007$  & $7.858\times10^{-15}$ & 41.51 & 16.27 & -2.099 &  -    & 13.982 & 43.95 & absorption   \\
1778  & J005929.7-361113 &  $0.0115\pm0.0003$  & $2.571\times10^{-14}$ & 39.87 & 12.38 & -3.140 & 41.50 & 14.042 & 41.81 & emission     \\
3654  & J022416.7-050323 &  -                  & $7.173\times10^{-15}$ &  -    & 15.38 & -2.492 &  -    &  -     &  -    & featureless  \\
3716  & J022456.2-050801 &  $0.0837\pm0.0008$  & $2.769\times10^{-14}$ & 41.67 & 12.57 & -3.032 &  -    & 13.839 & 43.63 & absorption   \\
3773  & J022536.4-050012 &  $0.0530\pm0.0002$  & $1.940\times10^{-14}$ & 41.11 & 12.40 & -3.252 & 41.76 & 13.298 & 43.45 & emission     \\
3776  & J022537.8-050223 &  -                  & $1.489\times10^{-14}$ &  -    & 13.94 & -2.751 &  -    &  -     &  -    & featureless  \\
3777  & J022538.2-050806 &  $0.0607\pm0.0003$  & $3.862\times10^{-15}$ & 40.53 & 15.67 & -2.645 & 41.17 & 15.670 & 42.61 & emission     \\
3778  & J022538.3-050423 &  0.0000             & $1.157\times10^{-14}$ & STAR  & 15.06 & -2.413 &  -    &  -     &  -    & STAR         \\
4009  & J023613.5-523036 &  $0.1117\pm0.0005$  & $8.520\times10^{-15}$ & 41.42 & 14.47 & -2.784 & 41.99 & 14.869 & 43.47 & emission     \\
4298  & J030927.5-765223 &  0.0000             & $1.597\times10^{-14}$ & STAR  & 14.99 & -2.301 &  -    &  -     &  -    & STAR         \\
4379  & J031256.5-765039 &  0.0000             & $1.104\times10^{-14}$ & STAR  & 13.12 & -3.209 &  -    &  -     &  -    & STAR         \\
4481  & J031723.1-442056 &  0.0000             & $2.835\times10^{-14}$ & STAR  & 14.82 & -2.119 &  -    &  -     &  -    & STAR         \\
4736  & J031829.8-441140 &  $0.0732\pm0.0005$  & $5.158\times10^{-14}$ & 41.82 & 12.19 & -2.914 & 42.25 & 14.366 & 43.30 & emission     \\
4765  & J031845.0-441042 &  $0.0735\pm0.0006$  & $8.163\times10^{-15}$ & 41.02 & 12.95 & -3.408 & 42.20 & 14.664 & 43.18 & emission     \\
4927  & J033831.4-351421 &  $0.139\pm0.003$    & $9.464\times10^{-14}$ & 42.66 & 13.31 & -2.202 & 40.85 &  -     &  -    & composite    \\
5570  & J043306.5-610760 &  $0.0589\pm0.0007$  & $1.304\times10^{-14}$ & 41.02 & 13.80 & -2.865 &  -    & 14.254 & 43.15 & absorption   \\
5663  & J043333.5-612427 &  $0.0604\pm0.0012$  & $1.382\times10^{-14}$ & 41.04 & 14.58 & -2.527 &  -    & 14.195 & 43.17 & absorption   \\
7355  & J055940.7-503218 &  $0.197\pm0.020$    & $8.546\times10^{-15}$ & 42.02 & 15.35 & -2.430 &  -    & 14.702 & 44.10 & absorption   \\
7400  & J060014.9-502230 &  -                  & $7.370\times10^{-15}$ &  -    & 15.56 & -2.411 &  -    &  -     &  -    & featureless  \\
9062  & J124238.5-111919 &  -                  & $1.940\times10^{-14}$ &  -    & 13.45 & -2.832 &  -    &  -     &  -    & not observed \\
9064  & J124239.1-112822 &  -                  & $8.434\times10^{-15}$ &  -    & 13.59 & -3.140 &  -    &  -     &  -    & not observed \\
9337  & J125204.5-292029 &  -                  & $3.552\times10^{-14}$ &  -    & 13.81 & -2.426 &  -    &  -     &  -    & not observed \\
9480  & J125718.4-171441 &  -                  & $1.182\times10^{-14}$ &  -    & 12.21 & -3.543 &  -    &  -     &  -    & not observed \\
9539  & J125819.1-171837 &  -                  & $1.512\times10^{-14}$ &  -    & 12.12 & -3.472 &  -    &  -     &  -    & not observed \\
9730  & J133032.4-014735 &  -                  & $5.251\times10^{-15}$ &  -    & 14.05 & -3.162 &  -    &  -     &  -    & not observed \\
9966  & J133527.7-342630 &  -                  & $1.014\times10^{-14}$ &  -    & 15.15 & -2.434 &  -    &  -     &  -    & not observed \\
12281 & J201329.7-414737 &  $0.1293\pm0.0006$  & $1.078\times10^{-14}$ & 41.67 & 14.69 & -2.591 & 43.49 & 14.047 & 43.93 & emission     \\
12308 & J201345.0-563713 &  $0.0541\pm0.0005$  & $1.114\times10^{-14}$ & 40.89 & 13.49 & -3.057 &  -    & 12.947 & 43.61 & absorption   \\
13137 & J213758.7-143611 &  $0.0524\pm0.0002$  & $4.446\times10^{-15}$ & 40.46 & 14.08 & -3.222 & 41.84 & 13.829 & 43.22 & emission     \\
14252 & J221543.0-173958 &  -                  & $2.929\times10^{-15}$ &  -    & 17.44 & -2.057 &  -    &  -     &  -    & not observed \\
14402 & J221726.0-082531 &  $0.0845\pm0.0002$  & $2.733\times10^{-14}$ & 41.67 & 13.97 & -2.477 &  -    & 13.759 & 43.67 & absorption   \\
14541 & J222110.0-244749 &  -                  & $8.297\times10^{-15}$ &  -    & 16.06 & -2.157 &  -    &  -     &  -    & featureless  \\
14608 & J222804.4-051751 &  -                  & $5.319\times10^{-15}$ &  -    & 15.03 & -2.762 &  -    &  -     &  -    & featureless  \\
14640 & J222818.3-050745 &  -                  & $7.741\times10^{-15}$ &  -    & 15.44 & -2.435 &  -    &  -     &  -    & not observed \\
14683 & J222834.0-052818 &  -                  & $9.300\times10^{-15}$ &  -    & 13.88 & -2.980 &  -    &  -     &  -    & featureless  \\
15022 & J225149.3-175225 &  $0.0776\pm0.0005$  & $1.711\times10^{-14}$ & 41.39 & 14.12 & -2.621 & 42.59 & 13.744 & 43.59 & emission     \\
15393 & J231421.6-424559 &  $0.0672\pm0.0006$  & $2.039\times10^{-14}$ & 41.35 & 12.47 & -3.203 &  -    & 13.391 & 43.63 & absorption   \\
15584 & J231851.8-423114 & $0.11432\pm0.00015$ & $1.460\times10^{-14}$ & 41.68 & 13.74 & -2.840 &  -    & 13.985 & 43.85 & absorption   \\
15631 & J232454.9-120459 &  $0.0809\pm0.0005$  & $1.164\times10^{-14}$ & 41.27 & 15.81 & -2.110 & 40.30 &  -     &  -    & composite    \\
15867 & J235340.6-102420 &  $0.0742\pm0.0011$  & $1.972\times10^{-14}$ & 41.42 & 12.15 & -3.347 &  -    & 13.985 & 43.47 & absorption   \\
15892 & J235405.7-101829 &  $0.0739\pm0.0006$  & $7.687\times10^{-15}$ & 40.99 & 13.38 & -3.262 &  -    & 13.780 & 43.53 & absorption   \\
15926 & J235418.1-102013 &  $0.0804\pm0.0003$  & $1.349\times10^{-14}$ & 41.32 & 14.48 & -2.580 & 43.38 & 14.040 & 43.51 & emission     \\
15997 & J235629.1-343743 &  $0.0523\pm0.0009$  & $1.437\times10^{-14}$ & 40.96 & 12.12 & -3.494 &  -    & 13.136 & 43.49 & absorption   \\
\hline
\end{tabular}
\end{table*}


\begin{thebibliography}{}
\scriptsize{
  \bibitem[Alexander et al.(2002)]{Alexander2002} Alexander, D. M., Aussel, H., Bauer, F. E., et al., 2002, ApJ, 568, L85
  \bibitem[Bauer et al.(2004)]{Bauer2004} Bauer, F. E., Alexander, D. M., Brandt, W. N., et al., 2004, AJ, 128, 2048
  \bibitem[Bell \& de Jong(2001)]{Bell2001} Bell, E. F., de Jong, R. S., 2001, ApJ, 550, 212
  \bibitem[Bondi et al.(2001)]{Bondi2001} Bondi, M., March\~{a}, M. J. M., Dallacasa, D., Stranghellini, C., 2001, MNRAS, 325, 1109
  \bibitem[Brocklehurst(1971)]{Brocklehurst1971} Brocklehurst, M., 1971, MNRAS, 153, 471
  \bibitem[Colbert et al.(2004)]{Colbert2004} Colbert, E. J. M., Heckman, T. M., Ptak, A. F., Strickland, D. K., 2004, ApJ, 602, 231
  \bibitem[David et al.(1992)David, Jones \& Forman]{David1992} David, L. P., Jones, C., Forman, W., 1992, ApJ, 388, 82
  \bibitem[Doane et al.(2004)]{Doane2004} Doane, N. E., Sanders, W. T., Wilcots, E. M., Juda, M., 2004, AJ, 128, 2712
  \bibitem[Fabbiano \& Trinchieri(1985)]{Fabbiano1985} Fabbiano, G., Trinchieri, G., 1985, ApJ, 196, 430
  \bibitem[Fabbiano \& Shapley(2002)]{Fabbiano2002} Fabbiano, G., Shapley, A., 2002, ApJ, 565, 908
  \bibitem[Fabbiano et al.(1987)]{Fabbiano1987} Fabbiano, G., Klein, U., Trinchieri, G., Wielebinski, R., 1987, ApJ, 111, 121
  \bibitem[Fabbiano et al.(1988)Fabbiano, Gioia \& Trinchieri]{Fabbiano1988} Fabbiano, G., Gioia, I. M., Trinchieri, G., 1988, ApJ, 324, 749
  \bibitem[Falomo(1996)]{Falomo1996} Falomo, R., 1996, MNRAS, 283, 241
  \bibitem[Forman et al.(1979)]{Forman1979} Forman, W., Schwarz, J., Jones, C., Liller, W., Fabian A. C., 1979, ApJ, 234, L27
  \bibitem[Georgakakis et al(2007)]{Georgakakis2007} Georgakakis, A., Rowan-Robinson, M., Babbedge, T. S. R., Georgantopoulos, I., 2007, MNRAS, 377, 203
  \bibitem[Georgakakis et al.(2006)]{Georgakakis2006} Georgakakis, A. E., Chavushyan, V., Plionis, M., Georgantopoulos, I., Koulouridis, E., Leonidaki, I., Mercado, A., 2006, MNRAS, 367, 1017
  \bibitem[Georgakakis et al.(2004)]{Georgakakis2004} Georgakakis, A. E., Georgantopoulos, I., Basilakos, S., Plionis, M., Kolokotronis, V., 2004, MNRAS, 354, 123
  \bibitem[Georgakakis et al.(2003)]{Georgakakis2003} Georgakakis, A., Georgantopoulos, I., Stewart, G. C., Shanks, T., Boyle, B. J., 2003, MNRAS, 344, 161
  \bibitem[Georgantopoulos et al.(2005)Georgantopoulos, Georgakakis \& Koulouridis]{Georgantopoulos2005} Georgantopoulos, I., Georgakakis, A., Koulouridis, E., 2005, MNRAS, 360, 782
  \bibitem[Gilfanov(2004)]{Gilfanov2004b} Gilfanov, M., 2004, MNRAS, 349, 146
  \bibitem[Gilfanov et al.(2004)Gilfanov, Grimm \& Sunayev]{Gilfanov2004} Gilfanov, M., Grimm, H.-J., Sunayev, R., 2004, MNRAS, 351, 1365
  \bibitem[Griffiths \& Padovani(1990)]{Griffiths1990} Griffiths, R. E., Padovani, P., 1990, ApJ, 360, 483
  \bibitem[Grimm et al.(2003)Grimm, Gilfanov \& Sunyaev]{Grimm2003} Grimm, H.-J., Gilfanov, M., Sunyaev, R., 2003, MNRAS, 339, 793
  \bibitem[Helou et al.(1985)Helou, Soifer \& Rowan-Robinson]{Helou1985} Helou, G., Soifer, B. T., Rowan-Robinson, M., 1985, ApJ, 298, L7
  \bibitem[Ho et al.(1997)Ho, Filippenko \& Sargent]{Ho1997} Ho, L. C., Filippenko, A. V., Sargent, W. L. W., 1997, ApJ, 112, 315
  \bibitem[Hornschemeier et al.(2005)]{Hornschemeier2005} Hornschemeier, A. E., Heckman, T. E., Ptak, A. F., Tremonti, C. A., Colbert, E. J. M., 2005, AJ, 129, 86
  \bibitem[Hornschemeier et al.(2003)]{Hornschemeier2003} Hornschemeier, A. E., Bauer, F. E., Alexander, D. M., et al., 2003, ApJ, 126, 575
  \bibitem[Irwin et al.(1994) Irwin, Maddox \& McMahon]{Irwin1994}  Irwin, M., Maddox, S., McMahon, R. G., 1994, Spectrum, 2, 14
  \bibitem[Isobe et al.(1990)]{Isobe1990} Isobe, T., Feigelson, E. D., Akritas, M. G., Babu, G. J., 1990, ApJ, 364, 104
  \bibitem[Kennicutt \& Kent(1983)]{Kennicutt1983} Kennicutt, R. C., Kent, S. M., 1983, AJ, 88, 1094
  \bibitem[Kennicutt et al.(1994)Kennicutt, Tamblyn \& Congdon]{Kennicutt1994} Kennicutt, R. C. Jr., Tamblyn, P., Congdon, C. E., et al., 1994, ApJ, 435, 22
  \bibitem[Kewley et al.(2002)]{Kewley2002} Kewley, L. J., Geller, M. J., Jansen, R. A., Dopita, M. A., 2002, AJ, 124, 3135
  \bibitem[Lasker et al.(1996)]{Lasker1996} Lasker, B. M., Doggett, J., McLean, B., et al., 1996, ASP Conf. Ser., 101, 88
  \bibitem[Lehmann et al.(2001)]{Lehmann2001} Lehmann, I., Hasinger, G., Schmidt, M., et al.,  2001, A\&A, 371, 833
  \bibitem[Londish et al.(2002)]{Londish2002} Londish, D., Croom, S. M., Boyle, B. J., et al., 2002, MNRAS, 334, 941
  \bibitem[Monet et al.(2003)]{Monet2003} Monet, D. G., Levine, S. E., Canzian, B., et al., 2003, ApJ, 125, 984
  \bibitem[Moshir et al.(1990)]{Moshir1990} Moshir, M., Kopan, G., Cornow, T., et al., 1990, IRAS Faint Source Catalog (ver. 2.0; Greenbelt; NASA/GSFC)
  \bibitem[Norman et al.(2004)]{Norman2004} Norman, C., Ptak, A., Hornschemeier, A., et al., 2004, ApJ, 607, 721
  \bibitem[Owen \& Warwick(2009)]{Owen2009} Owen, R. A., Warwick, R. S., 2009, MNRAS, in press \texttt{[ArXiv:astro-ph/0901.4263]}
  \bibitem[Persic \& Rephaeli(2007)]{Persic2007} Persic, M., Rephaeli, Y., 2007, A\&A, 463, 481
  \bibitem[Persic et al.(2004)]{Persic2004} Persic, M., Rephaeli, Y., Braito, V., Cappi, M., Della Ceca, R., Franceschini, A., Gruber, D. E., 2004, A\&A, 419, 849
  \bibitem[Ptak et al.(2007)]{Ptak2007} Ptak, A., Mobasher, B., Hornschemeier, A., Bauer, F., Norman, C., 2007, ApJ, 667, 826
  \bibitem[Ranalli et al.(2003) Ranalli, Comastri \& Setti]{Ranalli2003} Ranalli, P., Comastri, A., Setti, G., 2003, A\&A, 399, 39
  \bibitem[Read \& Ponman(2001)]{Read2001} Read, A. M., Ponman, T. J., 2001, MNRAS, 328, 127
  \bibitem[Read et al.(1997)Read, Ponman \& Strickland]{Read1997} Read, A. M., Ponman, T. J., Strickland, D. K., 1997, MNRAS, 286, 626
  \bibitem[Shapley et al.(2001)Shapley, Fabbiano \& Eskridge]{Shapley2001} Shapley, A., Fabbiano, G., Eskridge, P. B., 2001, ApJS, 137, 139
  \bibitem[Skrutskie et al.(2006)]{Skrutskie2006} Skrutskie, M. F., Cutri, R. M., Stiening, R., et al., 2006, AJ, 131, 1163
  \bibitem[Rosa-Gonz\`{a}lez et al.(2002)Rosa-Gonz\`{a}lez, Terlevich \& Terlevich]{RosaGonzalez2002} Rosa-Gonz\`{a}lez, D., Terlevich, E., Terlevich, R., 2002, MNRAS, 332, 283
  \bibitem[Savage \& Mathis(1979)]{Savage1979} Savage, B. D., Mathis, J. S., 1979, ARA\&A, 17, 73
  \bibitem[Sbarufatti et al.(2006)]{Sbarufatti2006} Sbarufatti, B., Treves, A., Falomo, R., Heidt, J., Kotilainen, J., Scarpa, R., 2006, AJ, 132, 1
  \bibitem[Stocke et al.(1991)]{Stocke1991} Stocke, J. T., Morris, S. L., Gioia, I. M., Maccacaro, T., Schild, R., Wolter, A., Fleming, T. A., Henry, J. P., 1991, ApJS, 76, 813
  \bibitem[Tajer et al.(2005)]{Tajer2005} Tajer, M., Trinchieri, G., Wolter, A., Campana, S., Moretti, A., Tagliaferri, G., 2005, A\&A, 435, 799
  \bibitem[Trinchieri \& Fabbiano(1985)]{Trinchieri1985} Trinchieri, G., Fabbiano, G., 1985, ApJ, 296, 447
  \bibitem[Trinchieri et al.(1989)Trinchieri, Fabbiano \& Bandiera]{Trinchieri1989} Trinchieri, G., Fabbiano, G., Bandiera, B., 1989, ApJ, 342, 759
  \bibitem[Troitsky(2008)]{Troitsky2008} Troitsky, S., 2008, MNRAS, 388L, 79
  \bibitem[Tyler et al.(2004)]{Tyler2004} Tyler, K., Quillen, A. C., LaPage, A., Rieke, G. H., 2004, ApJ, 610, 213
  \bibitem[Tzanavaris et al.(2006)Tzanavaris, Georgantopoulos \& Georgakakis]{Tzanavaris2006} Tzanavaris, P., Georgantopoulos, I., Georgakakis, A., 2006, A\&A, 454, 447
  \bibitem[Tzanavaris \& Georgantopoulos(2008)]{Tzanavaris2008} Tzanavaris, P., Georgantopoulos, I., 2008, A\&A, 480, 663
  \bibitem[Veilleux \& Osterbrock(1987)]{Veilleux1987} Veilleux, S., Osterbrock, D. E., 1987, ApJS, 63, 295
  \bibitem[Vogler et al.(1997)Vogler, Pietsch and Bertoldi]{Vogler1997} Vogler, A., Pietsch, W., Bertoldi, F., 1997, A\&A, 318, 768
  \bibitem[Warwick et al.(2007)]{Warwick2007} Warwick, R. S., Jenkins, L. P., Read, A. M., Roberts, T. P., Owen, R. A., 2007, MNRAS, 376, 1611
  \bibitem[Watson et al.(2003)]{Watson2003} Watson, M. G., Pye, J. P., Denby, M., et al., 2003, AN, 324, 89
  \bibitem[Zezas(2000)]{Zezas2000} Zezas, A., 2000, PhD thesis, University of Leicester
}
\end{thebibliography}
\end{document}